\documentclass[11pt,a4paper]{article}
\pdfoutput=1
\usepackage{jheppub}
\usepackage{subfigure}

\usepackage[latin1]{inputenc}
\usepackage{latexsym,amsmath,amsfonts,amssymb,enumerate,graphicx,slashed}

\def\F{{{\cal F}}}
\def\C{{{\mathbb C}}}

\def\Z{{{\mathbb Z}}}

\def\CL{{\mathcal{L}}}

\def\CC{{\cal C}}
\def\CD{{\cal D}}
\def\CH{{\cal H}}
\def\CL{{\cal L}}
\def\CN{{\cal N}}

 \newcommand{\be}{\begin{equation}}
\newcommand{\ee}{\end{equation}}
 \newcommand{\bal}{\begin{align}}
 \newcommand{\eal}{\end{align}}
\newcommand{\ben}{\begin{equation*}}
\newcommand{\een}{\end{equation*}}
\newcommand{\bea}{\begin{eqnarray}}
\newcommand{\eea}{\end{eqnarray}}
\newcommand{\bean}{\begin{eqnarray*}}
\newcommand{\eean}{\end{eqnarray*}}
\newcommand{\bes}{\begin{subequations}}
\newcommand{\ees}{\end{subequations}}

\newcommand{\zb}{\bar{z}}

\def\Imag{{\rm Im}\,}

\def\Tr{{\rm Tr}}

\def\eps{{\epsilon}}

\def\RR{{\mathbb R}}

\def\half{\frac{1}{2}}

\newcommand{\ket}[1]{|{#1}\rangle}

\newcommand{\bra}[1]{\langle{#1}|}

\newcommand{\ev}[1]{\langle{#1}\rangle}
\newcommand{\bev}[1]{\left\langle{#1}\right\rangle}
\DeclareMathOperator{\tr}{tr}
\DeclareMathOperator{\sgn}{sgn}
\DeclareMathOperator{\RealPart}{Re}

\title{Bose-Fermi duality and entanglement entropies}

\author{Matthew Headrick${}^{1,2}$, Albion Lawrence${}^1$, and Matthew Roberts${}^3$} 

\affiliation{${}^1$ Martin Fisher School of Physics, Brandeis University, Waltham, Massachusetts, USA}

\affiliation{${}^2$ Center for the Fundamental Laws of Nature, Harvard University, Cambridge, Massachusetts, USA}

\affiliation{${}^3$ Department of Physics and Center for Cosmology and Particle Physics, New York University, New York, New York, USA}

\abstract{Entanglement (R\'enyi) entropies of spatial regions are a useful tool for characterizing the ground states of quantum field theories. In this paper we investigate the extent to which these are universal quantities for a given theory, and to which they distinguish different theories, by comparing the entanglement spectra of the massless Dirac fermion and the compact free boson in two dimensions. We show that the calculation of R\'enyi entropies via the replica trick for any orbifold theory includes a sum over orbifold twists on all cycles. In a modular-invariant theory of fermions, this amounts to a sum over spin structures. The result is that the R\'enyi entropies respect the standard Bose-Fermi duality. Next, we investigate the entanglement spectrum for the Dirac fermion \emph{without} a sum over spin structures, and for the compact boson at the self-dual radius.  These are not equivalent theories; nonetheless, we find that (1) their second R\'enyi entropies agree for any number of intervals, (2) their full entanglement spectra agree for two intervals, and (3) the spectrum generically disagrees otherwise. These results follow from the equality of the partition functions of the two theories on any Riemann surface with imaginary period matrix. We also exhibit a map between the operators of the theories that preserves scaling dimensions (but not spins), as well as OPEs and correlators of operators placed on the real line. All of these coincidences can be traced to the fact that the momentum lattice for the bosonized fermion is related to that of the self-dual boson by a 45$^\circ$ rotation that mixes left- and right-movers.
}

\arxivnumber{1209.2428}

\preprint{BRX-TH-656, NSF-KITP-12-096}

\begin{document}

\maketitle

\section{Introduction}

The quantum entanglement between spatial regions, as quantified by entanglement entropies and entanglement R\'enyi entropies, is an important tool for characterizing the infrared behavior of extended quantum systems. In theories with a mass gap, these quantities provide a characterization of topological phases where no local order parameter exists \cite{Hamma200522,Kitaev:2005dm,PhysRevLett.96.110405,PhysRevLett.103.261601}. When the infrared limit is a nontrivial two-dimensional conformal field theory, computing the entropy of an interval (for example in a lattice model) provides an efficient way to determine its central charge \cite{Calabrese:2004eu}. The entropies of more than one interval in a CFT depend on the full operator spectrum, and therefore give more refined information about the theory; as an example, for a free compact boson the R\'enyi entropies of two intervals depend on the compactification radius \cite{Calabrese:2009ez,Calabrese:2009qy}.

An important question is thus whether the entanglement entropies (by which we mean both von Neumann and R\'enyi entropies) of spatial regions in the ground state of a field theory characterize that theory, in the following precise senses:
\begin{enumerate}
\item The entanglement entropies should be the same regardless of the presentation of the theory (up to non-universal cutoff-dependent terms).  That is, the same theory could have two different Lagrangian descriptions, but the underlying spectra of states and local operators are the same; in this case, for a quantity to be universal it should give the same answer for both presentations. For example, in the case of the compact boson, the R\'enyi entropies are T-duality invariant \cite{Calabrese:2009ez,Calabrese:2009qy}.
\item The entanglement entropies should distinguish different theories.
\end{enumerate}
Surprisingly, these statements appear to be challenged already by some very simple quantum field theories, namely the free massless Dirac fermion and the compact free boson in two dimensions. In the case of the Dirac fermion, the R\'enyi entropies for any number of intervals have been computed by Casini, Fosco, and Huerta \cite{Casini:2005rm,Casini:2009vk,Casini:2009sr}. In the case of the compact boson, the R\'enyi entropies for two intervals have been computed by Calabrese, Cardy, and Tonni \cite{Calabrese:2009ez,Calabrese:2009qy}. Since the boson at radius $R=\sqrt{2}R_{\rm sd}$ (where $R_{\rm sd}$ is the self-dual radius) is known to be dual to a theory of a Dirac fermion, one might expect that the entropies computed by CFH would agree with those computed by CCT at that value of $R$. In fact, they do not, seeming to violate point (1) above. Various explanations have been put forward for this discrepancy, such as that the bosonization relating the two theories is a non-local transformation on the fields appearing in the path integral, and hence might not preserve the factorization of the Hilbert space according to spatial regions \cite{Casini:2009sr,Calabrese:2009qy}. Even more curiously, the Dirac fermion R\'enyi entropies \emph{do} agree with those for the boson at $R=R_{\rm sd}$, despite the fact that these two theories are certainly not dual to each other. By conformal invariance the R\'enyi entropies for two intervals are effectively functions of the cross-ratio of the four endpoints, so this is a non-trivial agreement between an infinite number of functions of one real variable. This coincidence would seem to threaten point (2).

In this paper we will show that in fact both points continue to hold. The key to resolving the first discrepancy is to recall that there are two versions of the free Dirac fermion theory. We will show in Section \ref{sec:ent_entropy} that the calculation of CFH \cite{Casini:2005rm}, as well as a subsequent calculation by Casini and Huerta giving the same result by a different method \cite{Casini:2009vk}, specifically produces the entanglement entropies for the Dirac fermion theory without any projection on fermion number (and containing only NS-NS sector operators). This is not a modular-invariant theory, and is not dual to the boson at radius $\sqrt{2} R_{\rm sd}$. Instead, it is only after a certain $\Z_2$ gauging, which introduces R-R operators and removes the fermionic ones, that the theory becomes modular-invariant and dual to the $\sqrt{2} R_{\rm sd}$ boson. We will show that when the R\'enyi entropies of theories with such discrete gaugings (including orbifold theories) are calculated in terms of the partition functions on Riemann surfaces, one must perform a sum over all twists by the gauge symmetry.  This guarantees that the R\'enyi entropies for arbitrary numbers of intervals are invariant under Bose-Fermi duality. Thus, point (1) above is satisfied in this case.

Having dispensed with the boson at $R=\sqrt{2}R_{\rm sd}$, in the remaining sections we will explore the relationship between the original (unprojected) Dirac fermion and the self-dual boson. Our goals will be to understand the origin of the surprising agreement between their R\'enyi entropies for two intervals, to discover whether it extends to more than two intervals, and to see whether such a coincidence could happen in other (perhaps more complicated) theories. In doing so we will find that the theories are related by a new kind of quasi-duality, which we call a ``real duality", that goes well beyond R\'enyi entropies.

In Section \ref{sec:renyi_entropies} we will study the partition functions of the theories on Riemann surfaces, finding that they agree precisely when its period matrix is imaginary. Using the symmetries of the Riemann surface involved in computing the $n$th R\'enyi entropy for $N$ intervals, we show that this condition holds when $N=2$ for any $n$ (explaining the agreement found before), and also for any $N$ when $n=2$, but not more generally; the R\'enyis for $n>2$, $N>2$ do indeed distinguish between these two theories, so condition (2) above is satisfied. (Therefore the von Neumann entropies, which are related to the R\'enyi entropies by an analytic continuation in $n$, also presumably distinguish between the theories for $N>2$.) Since the R\'enyi entropies are known for the Dirac fermion for all $N$ \cite{Casini:2009sr}, as a bonus of our analysis we learn what the $n=2$ R\'enyis are for all $N$ (see equation \eqref{CasiniHuerta}).

The agreement between the partition functions for imaginary period matrices is due to the following relationship between the theories: If we bosonize the Dirac fermion, its momentum lattice (which is simply $\Z^2$) is related to the one for the self-dual boson by a 45$^\circ$ rotation (see figure \ref{momentum_lattice}). This rotation preserves the scaling dimensions of the corresponding momentum operators, but, since it mixes left- and right-movers, it changes their spins. Since the two theories also have the same oscillator structure, they have the same total spectrum of scaling dimensions, and hence the same torus partition function for imaginary $\tau$, i.e.\ on a rectangular torus. The agreement for higher-genus Riemann surfaces with imaginary period matrices is a generalization of this statement.

In Section \ref{sec:real_duality}, we will use the 45$^\circ$ rotation on the momentum lattices to define a canonical one-to-one correspondence between the operators of the two theories that preserves not only the scaling dimensions, but also (1) the OPE of any two operators that are separated by a real interval; (2) the correlator on the plane of arbitrary operators with positions on the real axis; and (3) the action of the mixed Virasoro generators $L_n+\tilde L_n$. We refer to this relationship between the two theories as a ``real duality". The coincidence of correlators is directly related to the statement about Riemann surfaces with imaginary period matrices, assuming a certain conjecture about their Schottky parameters.

In Section \ref{sec:discussion}, we extend our results on the agreements and disagreements between the boson and fermion theories to finite temperature and finite volume, including comparing our results to those found when calculating the entanglement negativity \cite{Calabrese:2012ew,Calabrese:2012nk}, and we discuss generalizations to other pairs of theories. We then discuss various larger issues connected to our work, returning in particular to the question we started with, whether entanglement entropies characterize theories.

There are also three appendices.  Appendix \ref{app:resummation} is a derivation of a specific expression for the partition function of the self-dual boson with radius $R = 1$, filling in a calculation needed in section \ref{sec:renyi_entropies}.  Appendix \ref{app:dihedral_rep} summarizes the irreducible representations of the dihedral group $D_n$, also needed for section \ref{sec:renyi_entropies}.  Finally, Appendix \ref{app:genl_boson_renyi} presents a rederivation of the results of \cite{Calabrese:2009qy}\ for the R\'enyi entropies for two intervals of a boson at any radius, and extends this result to the second R\'enyi entropy for any number of intervals.


\section{Entanglement entropies and discrete gauge symmetries for $1+1$ CFTs}\label{sec:ent_entropy}

In this section we will revisit a puzzle that arises in the computation of the entanglement entropies for the free massless fermion and free massless boson in $1+1$ dimensions.  We will begin in subsection \ref{subsec:defns}, with a  review of the definition of entanglement entropies in conformal field theory.  In subsection \ref{subsec:results_free_cfts}, we will review recent results for the free Dirac fermion and the boson in $1+1$ dimensions, which seem to indicate that the R\'enyi entropies are not invariant under Bose-Fermi duality \cite{Casini:2009sr,Calabrese:2009qy}.  To prepare for a deeper investigation of this question, we will review the ``replica trick" calculation of entanglement entropies in subsection \ref{subsec:replica_trick}.  In subsection \ref{subsec:disc_gauge_sym} we will extend that prescription to theories with discrete abelian gauge symmetries, which includes both bosonic orbifolds and modular-invariant fermionic CFTs.  The invariance of the R\'enyi entropies under bosonization will follow automatically from the results of this subsection and old results about partition functions and bosonization \cite{AlvarezGaume:1986es,AlvarezGaume:1987vm}. Finally, in subsection \ref{subsec:fermion_bc} we will resolve the puzzle by showing that the fermionic theory in question is not in fact dual to the free boson in question.

The results we will review in subsection \ref{subsec:replica_trick} also include an equality between the entanglement R\'enyi entropies for two theories that are \emph{not} dual to each other. Section \ref{sec:renyi_entropies} of this paper will be devoted to explaining this puzzling equality.

\subsection{Definition of entanglement entropies}\label{subsec:defns}

We will consider $1+1$-dimensional conformal field theories (CFTs) on the plane.  For thorough reviews of the entanglement entropies in these theories, we recommend \cite{Casini:2009sr,Calabrese:2009qy}.

\begin{figure}[thbp]
\begin{center}
\includegraphics[scale=.9]{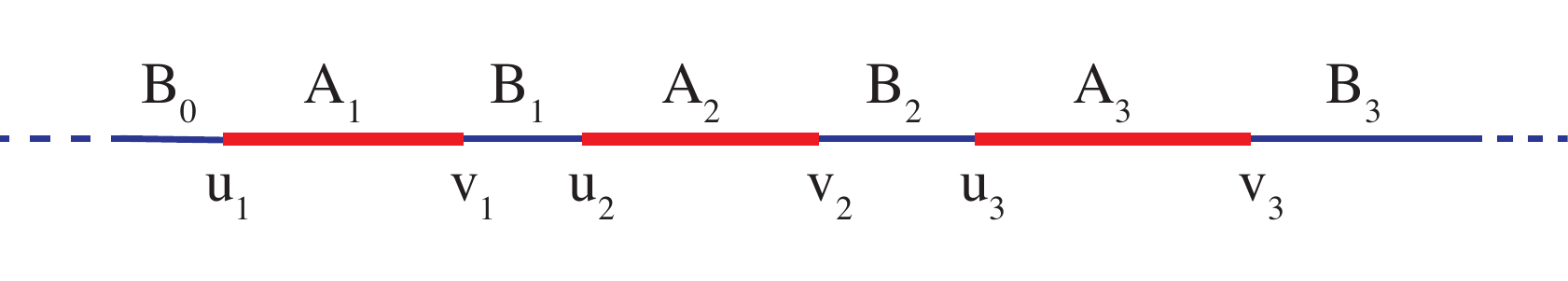}
\caption{A line divided up into consecutive intervals.  We will be considering the density matrix for the field theory on the intervals $A_{i = 1,2,3}$ upon tracing out the local degrees of freedom in the intervals $B_{i = 0,1,2,3}$.}
\label{Intervals}
\end{center}
\end{figure}

Let $A$ be the union of $N$ intervals on the real axis, $A=\cup_{i=1}^NA_i$, $A_i=[u_i,v_i]$ ($u_1<v_1<u_2<\cdots$). We denote the complement as $B = \cup_{i = 0}^N B_i$, where $B_i = [v_i,u_{i+1}]$, with $v_0 = - \infty$, $u_{N+1} = \infty$. (See Fig.\ \ref{Intervals}.) We will assume that for a local theory, we can decompose the Hilbert space as $\CH = \CH_A \otimes \CH_B$, where $\CH_{A,B}$ is the Hilbert space of degrees of freedom localized on $A,B$.  In practice one must regularize the theory; if we put the theory on a lattice, then this decomposition should make sense.

The density matrix for the vacuum is $\rho=\ket{0}\bra{0}$; the reduced density matrix on $A$ is $\rho_A = \tr_{\CH_B}\rho$. Its R\'enyi entropy of order $n$, called the ``entanglement R\'enyi entropy" since it is a measure of the amount of entanglement between $\CH_A$ and $\CH_B$, is defined by
\begin{equation}
S_n(A) = \frac1{1-n}\ln\tr\rho_A^n\,,
\end{equation}
where $n$ is a positive real parameter not equal to 1. Typically the R\'enyi entropy is computed for integer values of $n$. Knowing $S_n(A)$ for all integer $n>1$ is then enough in principle to fix, by analytic continuation, the value for all positive real $n$. In particular, the value of the analytically continued function at $n=1$ is the entanglement (von Neumann) entropy $S(A) = -\tr\rho_A\ln\rho_A$. One can also consider various interesting linear combinations, such as the mutual R\'enyi information between two intervals: $I_n(A_1:A_2) = S_n(A_1) + S_n(A_2) - S_n(A_1\cup A_2)$ (or, more generally, between two disjoint sets of intervals). In any computation, the R\'enyi and von Neumann entropies will diverge.  The goal is then to extract universal, regulator-independent quantities.  For example, the divergent parts of the R\'enyi entropies cancel for the mutual R\'enyi information

One way to compute $S_n(A)$ is to find an explicit expression for the reduced density matrix $\rho_A$ in some basis, and from it directly compute $\tr\rho_A^n$. To our knowledge the only theory for which this has been accomplished is the free massless fermion \cite{Casini:2009vk,Casini:2009sr}. The more common method, which we will review in \S\ref{subsec:replica_trick}, is the so-called replica trick, in which $\tr\rho_A^n$ is expressed in terms of the Euclidean partition function on an $n$-sheeted Riemann surface with branch cuts along the intervals $A_i$.

\subsection{Results for free CFTs}\label{subsec:results_free_cfts}

The classic result by Holzhey, Larsen, and Wilczek, derived using the replica trick, is that the R\'enyi entropies for one interval are the same for all CFTs, up to an overall factor of the central charge \cite{Holzhey:1994we}:\footnote{There can also be an $A_1$-independent  finite term. Such terms are related to the UV cutoff scheme employed and cancel out of finite quantities like mutual R\'enyi informations, so we neglect them throughout this paper.}
\begin{equation}
S_n(A_1) = \frac c6\left(1+\frac1n\right)\ln\frac{v_1-u_1}\epsilon\,,
\end{equation}
where $\epsilon$ is an ultraviolet cutoff length.

On the other hand, for more than one interval the R\'enyi entropies depend on more than just the central charge. The only theory for which the entropies have been computed exactly for any number of intervals is the free Dirac fermion, which was accomplished both using the replica trick (by Casini, Fosco, and Huerta \cite{Casini:2005rm}) and by deriving an explicit formula for $\rho_A$ (by Casini and Huerta \cite{Casini:2009vk}).  The result is a remarkably simple formula, in which the $n$-dependence factors out entirely from the dependence on the configuration of intervals:\footnote{The function $\Xi$ can also be expressed in a couple of other useful ways: $\Xi(A) = -\ln|\det M|$, where $M$ is an $N\times N$ matrix with entries $M_{ij} = \epsilon/(v_j-u_i)$; and $\Xi(A) = \sum_i\ln((v_i-u_i)/\epsilon)+\sum_{i<j}\ln(1-x_{ij})$, where $x_{ij} = (v_i-u_i)(v_j-u_j)/(u_j-u_i)(v_j-v_i)$. From the second form one sees that the mutual R\'enyi information between two sets of intervals $A,B$ takes a particularly simple form, as the integral of a bilocal quantity: $I_n(A:B) = -\sum_{i\in A,j\in B}\ln(1-x_{ij}) = \int_Ads\int_Bds'(s-s')^{-2}$.}
\begin{multline}\label{CasiniHuerta}
S_n^{\rm (f)}(A) = \frac16\left(1+\frac1n\right)\Xi(A)\,,\\
\Xi(A) = \sum_{i,j}\ln|v_j-u_i| - \sum_{i<j}\ln(u_j-u_i) - \sum_{i<j}\ln(v_j-v_i) - N\ln\epsilon\,.
\end{multline}

Calabrese, Cardy, and Tonni \cite{Calabrese:2009ez} computed the R\'enyi entropies for two intervals for the compact boson at arbitrary radius $R$ using the replica trick. Their result is quite a bit more complicated than \eqref{CasiniHuerta}, but it can be conveniently written as \eqref{CasiniHuerta} plus a correction term:
\begin{equation}\label{CCT}
S_n^R(A) = S_n^{\rm (f)}(A) + \frac1{1-n}\ln\mathcal{F}^R_n(x)\,.
\end{equation}
The correction term is finite (does not involve $\epsilon$), and depends only on the conformally invariant cross-ratio of the four endpoints:
\begin{equation}\label{crossratio}
x = \frac{(v_1-u_1)(v_2-u_2)}{(u_2-u_1)(v_2-v_1)}\,.
\end{equation}
$\mathcal{F}^R_n$ is a ratio of Riemann-Siegel theta functions
\begin{equation}\label{Fdef}
\mathcal{F}^R_n(x) = \frac{\vartheta(0|\eta\Gamma)\vartheta(0|\Gamma/\eta)}{\vartheta(0|\Gamma)^2}\,,
\end{equation}
where $\eta = R^2/R_{\rm sd}^2$ ($R_{\rm sd}$ is the self-dual radius), $\Gamma$ is an $x$-dependent $(n-1)\times(n-1)$ matrix with the following entries:
\begin{equation}\label{Gammadef}
\Gamma_{rs} = \frac{2i}n\sum_{k=1}^n\sin\left(\frac{\pi k}n\right)\cos\left(\frac{2\pi k(r-s)}n\right)\frac{{}_2F_1(k/n,1-k/n;1,1-x)}{{}_2F_1(k/n,1-k/n;1,x)}\,,
\end{equation}
and the Riemann-Siegel theta function (at the origin) is
\begin{equation}
\vartheta(0|\Gamma) = \sum_{m\in\Z^{n-1}}e^{i\pi\Gamma_{rs}m^rm^s}\,.
\end{equation}

Although the expression for $S^R_n(A)$ is complicated, three key points are clear just from \eqref{Fdef}: 
\begin{itemize} 
\item $\mathcal{F}^R_n(x)$, and hence $S^R_n(A)$, are invariant under T-duality. So, at least in this case, the presentation of the theory does not affect the entanglement spectrum. 
\item At the self-dual radius ($\eta=1$), $\mathcal{F}^{R_{{\rm sd}}}_n(x)=1$ identically, hence $S_n^{R_{\rm sd}}(A)=S_n^{\rm(f)}(A)$ for any $n$ and any $A$. 
\item At $R=\sqrt{2}R_{{\rm sd}}$ ($\eta=2$), where the boson is dual to a theory of a Dirac fermion, $\mathcal{F}^R_n(x)\neq1$ (this can easily be confirmed numerically, in case the reader is worried about theta-function conspiracies), hence $S_n^{\sqrt{2}R_{\rm sd}}(A)\neq S_n^{\rm(f)}(A)$. 
\end{itemize}

The discrepancy between $S_n^{\sqrt{2}R_{\rm sd}}(A)$ and $S_n^{\rm(f)}(A)$ has been noted in the literature but has not been satisfactorily resolved.  Casini and Huerta propose that the mismatch is due to the fact that the bosonization transformation is non-local; therefore, although the two theories have the same Hilbert space, the way that that Hilbert space gets cut up according to spatial regions in the two presentations might be different \cite{Casini:2009sr}. Calabrese and Cardy imply that the discrepancy is related to the Lagrangian used in computing the entanglement entropies of the fermion \cite{Calabrese:2009qy}. This is not an unreasonable thing to expect---at the level of the path integral, bosonization is not a local transformation of the fields we integrate over. It is therefore fair to ask whether it is a local transformation at the level of the Hilbert space, that is, whether the factorization of the Hilbert space by spatial regions is invariant under arbitrary duality transformations.

However, interpreting the mismatch requires some care.  The correct Bose-Fermi equivalence is between the boson at radius $\sqrt{2} R_{\rm sd}$ and the Dirac fermion gauged in a specific way by the $\Z_2$ fermion number \cite{Elitzur:1986ye,AlvarezGaume:1986es}. For example, the fermionic theory without such a gauging is not a modular-invariant theory, while the free boson is.\footnote{Note that one may add additional ``topological terms" to the bosonic theory which spoil modular invariance and lead to a theory which is precisely equivalent to a fermionic theory with fixed spin structure \cite{AlvarezGaume:1987vm}. This cannot be the theory of the modular-invariant bosonic at $R = R_{{\rm sd}}$.}  After this gauging, the spectra and the algebras of local operators are identical.  Since this data defines a two-dimensional CFT, we might expect that computations of the position-space entanglement entropies should be the same whether computed in the bosonic or fermionic representation. More precisely, the scheme used for cutting off the theory may depend on the representation of the theory, but universal quantities such as the mutual information $I_n(A_1:A_2)$ should not.

In the remainder of this section we will explain this apparent mismatch.

\subsection{R\'enyi entropies via the replica trick}\label{subsec:replica_trick}

\begin{figure}[thbp]
\begin{center}
\includegraphics[scale=.5]{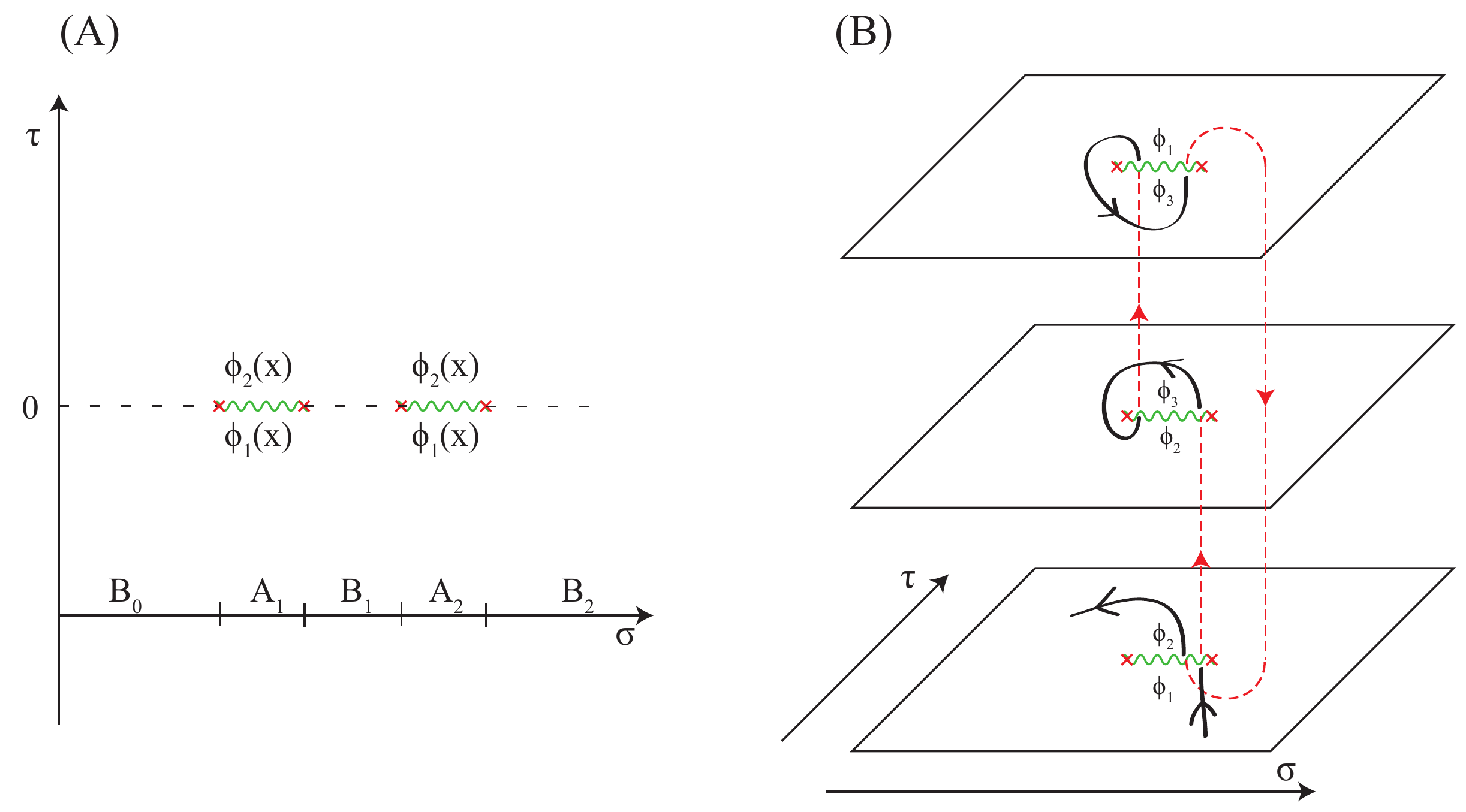}
\caption{A: The reduced density matrix $\rho(\phi_1,\phi_2)$  computed via a path integral on the complex plane with cuts $A_i$ on the real line, boundary conditions $\phi = \phi_1$ at the top of the cut and $\phi = \phi_2$ on the bottom of the cut. B:  The Riemann surface $\Sigma_{n,N}$ constructed as an $n$-fold branched cover of the complex plane, with branch cuts at $A_i$ glued together as shown.}
\label{cutplanetwo}
\end{center}
\end{figure}

Let us review the calculation of the density matrix $\rho$ and the R\'enyi entropies via path integrals in a two-dimensional conformal field theory $\CC$ (again, see also \cite{Calabrese:2009qy}). $\phi$ denotes all microscopic fields in the theory; a matrix element of $\rho$ is $\rho(\phi_1,\phi_2) = \langle \phi_1|0\rangle\langle 0 | \phi_2 \rangle$, where $|\phi_{1,2}\rangle$ are field eigenstates with eigenfunctions $\phi_{1,2}(x)$. These inner products can be represented via path integrals,
\be
\langle \phi_i | 0 \rangle = \CN\int\limits_{\phi(x,0) = \phi_i(x)}  \CD \phi \exp \left[  - \int_{ -\infty}^0d\tau \int_{-\infty}^{\infty}d\sigma~ \CL(\phi)\right]. \label{eigen_inner_product}
\ee
To trace out the spatial region $B$, let us use $\phi_{A,B}$ to denote the function $\phi(x \in (A,B), \tau)$. We decompose the boundary conditions in (\ref{eigen_inner_product}) into those at $x\in A$ and $x \in B$, so that
\be
\langle\phi_A,\phi_B | 0 \rangle =  \CN\int\limits_{\phi(x\in A,0) = \phi_A(x) \atop\phi(x\in B,0) = \phi_B(x)}  \CD \phi \exp \left[  - \int_{ -\infty}^0d\tau \int_{-\infty}^{\infty}d\sigma~ \CL(\phi)\right].\label{phi_A_B_vac}
\ee
Tracing out the region $B$ in the \emph{unorbifolded} theory is now simple: for functions $\phi_{1,2}$ on $A$,
\be
\rho_A(\phi_1,\phi_2) = \frac{1}{Z_1} \int \CD\phi_B \langle \phi_A=\phi_1,\phi_B|0\rangle \langle 0| \phi_A=\phi_2,\phi_B\rangle\,,\label{eq:unorb_red_density}
\ee
where $Z_1$ is the partition function of the CFT on the Riemann sphere. (See figure \ref{cutplanetwo}A.)
The R\'enyi entropies are:
\be\label{eq:fpirenyi}
\Tr \rho_A^n = \frac{1}{Z_1^n} \int \prod_{a=1}^N\CD \phi_{a}~\rho_A(\phi_{a},\phi_{a+1})\,,~ \phi_{N+1} = \phi_1.
\ee

The integral in (\ref{eq:fpirenyi}) can be done by ``replicating" the Euclidean spacetime.  The $k$th integrand in the product, $\rho_A(\phi_{a},\phi_{a+1})$, is the path integral on the complex plane, with cuts on the real line at $A_i$, and boundary conditions $\phi_a$ at the "bottom" of the cut and $\phi_{a+1}$ at the "top" of the cut.   In taking the product and integrating over all of the $\phi_a$s, we are taking $n$ copies of the plane and gluing them together in cyclic order by identifying the top of the cut on sheet $k=1,\ldots n-1$ with the bottom of the cut on sheet $k+1$, and glued the top of the cuts on sheet $n$ with the bottom of the cuts on sheet 1. (See figure \ref{cutplanetwo}B).  This is a singular Riemann surface $\Sigma_{n,N}$ with genus $g = (n-1)(N-1)$, described as an $n$-fold branched cover of the sphere over $N$ branch cuts.  The result is that 
\be
	\tr \rho_A^n = \frac{Z_{n,N}}{Z^n}
\ee
where $Z_{n,N}$ is the partition function of the CFT on $\Sigma_{n,N}$.

While we used a basis of field eigenstates to construct the theory, this is not necessary in principle.  We could have used any other basis that respects the decomposition $\CH = \CH_A \otimes \CH_B$.

\subsection{Including discrete gauge symmetries}\label{subsec:disc_gauge_sym}

Next, consider orbifolds of our theory $\CC$ by a discrete symmetry group $G$. For the sake of simplicity, let us consider the case of a bosonic orbifold $\CC/G$ with the unorbifolded CFT $\CC$ described as above by scalar fields $\phi$, and a $G$-action $\phi \mapsto g\phi$.  The $G$-action could be a finite rotation or a discrete translation.  We will restrict to the case that $G$ is abelian. Before tracing out any spatial regions, (\ref{phi_A_B_vac}) still holds.  

\begin{figure}[thbp]
\begin{center}
\includegraphics[scale=.45]{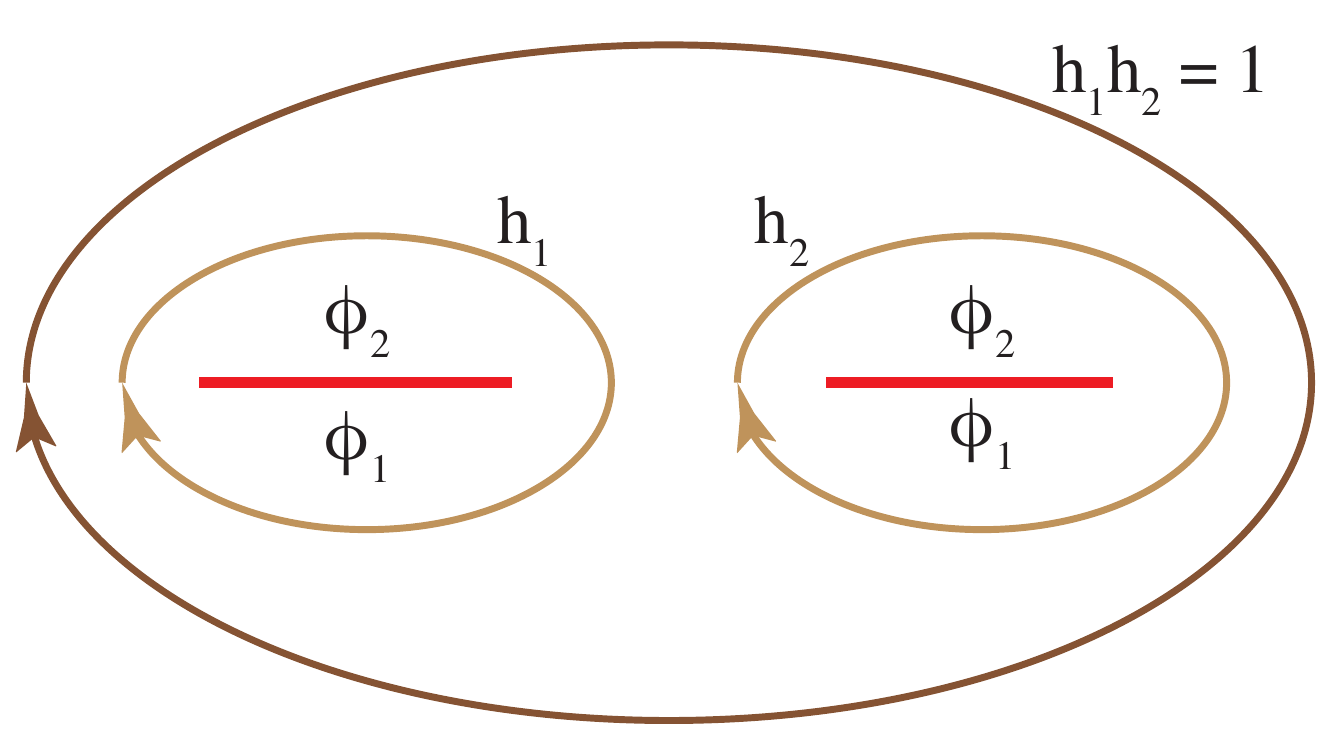}
\caption{The cut plane used to calculate the density matrix for 2 intervals.  The fields on each side of the slit are independent, as in (\ref{gauged_density_matrix}).  For R\'enyi entropies of the reduced density matrix corresponding to the CFT vacuum, the fields are untwisted for a circle which encloses all cuts $A_i$.  These can be deformed to the sum of the two loops shown which encircle the cuts.  The bosonic fields can be twisted about these loops, so long as the ordered product is the identity.
}
\label{cutcylinder}
\end{center}
\end{figure}

Breaking up $\phi_B$ into its values $\{\phi_{B_i}\}$ on each interval $B_i$, the reduced density matrix for the intervals $A_i$ in the orbifold theory is:
\be
\rho_A(\phi_1,\phi_2) = \frac{1}{Z_1}\sum_{\{g_i\}\in G^N} \int \CD\phi_{B_1} \cdots \CD\phi_{B_N} \langle \phi_A=\phi_1,\{g_i\phi_{B_i}\}|0\rangle \langle 0| \phi_A=\phi_2,\{g_i\phi_{B_i}\}\rangle\,.\label{gauged_density_matrix}
\ee
Here $g_i \in G$; in taking the trace over degrees of freedom in the intervals $B_i$, we have identified $\phi$ up to discrete  gauge transformations, so that the trace is being taken in $\CC/G$.  $Z_{1}$ is again just the path integral on the Riemann sphere, and its presence ensures that $\tr_A \rho_A = 1$.  The result is that the reduced density matrix is the sum over path integrals on the cut plane shown in figure \ref{cutcylinder}, with each element of the sum corresponding to twists of the field by $h_i = g_i g_{i-1} \in G$ as one transports the fields around the cuts $A_i$. 

The fields on either side of  \emph{all} of the cuts are still untwisted as one transports them from $-\infty$ to $\infty$ along a curve parallel to the real line: they can be deformed along the imaginary axis to $\pm i \infty$, and the system is in the vacuum which is generally an untwisted state. Thus, if we take the sum of a left-directed contour above the cuts  and a right-directed contour below the cuts, we can deform them so that they become the sum of single contours around each cut.  Since the fields have zero twist around the initial contours, the products of the twists around all of the cuts must be equal to the identity: $\prod_{i=1}^N h_i = \bf 1$. 

Next, we wish to compute the R\'enyi entropies, by computing
\be\label{eq:orbtrace}
	\tr \rho_N^n =  \sum_{\{g_i\}\in G^N} \int \prod_i d\phi(A_i) \rho(g_i \phi(A_i), \phi(A_{i+1}))
\ee
Once again we have sewed together fields $\phi_i$  along intervals $A_i$ up to identification by the gauge group $G$.  The result is
\be\label{eq:Renyi}
e^{(1-n)S_n(A)}\equiv	\tr \rho_N^n = \frac{Z_{n,N}}{Z_1^n}\,.
\ee
The expressions (\ref{gauged_density_matrix}),(\ref{eq:orbtrace}) make it clear that $Z_{n,N}$ is the partition function for the orbifold CFT on $\Sigma_{n,N}$, in which we have summed over all $G$ twists about all non-contractible cycles (again, the cycles at infinity on each sheet are taken to be trivial). In other words, one treats the Riemann surfaces exactly as one would Riemann surfaces for string perturbation theory for orbifold backgrounds.

We discussed bosonic orbifolds for simplicity, but this argument will be identical for gaugings of fermionic theories. When the fermion number itself is gauged, $Z_{n,N}$ will correspond to the sum over all spin structures of fermion partition functions on $\Sigma_{N,n}$. 

We are now in a position to argue that the entanglement entropies are invariant under Bose-Fermi duality. This duality is between the massless Dirac fermion gauged by $\Z_2$ fermion number and the free boson on a target space circle with radius $R = \sqrt{2} R_{{\rm sd}}$.  It is known that the equivalence holds for partition functions on any Riemann surface, so long as one correctly sums over the fermion spin structures \cite{Elitzur:1986ye,AlvarezGaume:1986es,AlvarezGaume:1987vm}. Since the R\'enyi entropies are determined by these partition functions, they are guaranteed to match. 

More generally, the results of this section allow one to discuss entanglement entropies for a large class of orbifold theories.  There are two more complicated generalizations which we leave for future work.  One is the case of orbifolds with discrete torsion, in which the different twisted partition functions on a Riemann surface are added with nontrivial phases. The second is the case of nonabelian orbifold groups.  We suspect that the result will be the same---one treats the Riemann surfaces in the R\'enyi entropy calculations precisely as one would the Riemann surfaces in string perturbation theory calculations (but without the integration over moduli).

\subsection{Boundary conditions in the Dirac fermion calculations}\label{subsec:fermion_bc}

\def\CT{{\cal T}}

Given the results of the prior section, a candidate explanation for the discrepancy described in \cite{Casini:2009sr} is that they are working with a different gauging or with the ungauged theory.  We will argue that the latter is in fact the case, by examining both the replica calculation \cite{Casini:2005rm,Casini:2009sr} and the direct construction of the reduced density matrix \cite{Casini:2009vk,Casini:2009sr}.

We begin with the calculation of $Z_{n,N}$ using the replica trick \cite{Casini:2005rm,Casini:2009sr}. In applying the replica trick to a theory containing fermionic fields, one needs to be careful about boundary conditions for the fermions. In a theory in which fermion number is gauged, it follows from the discussion in the previous subsection that one should sum over the partition functions with NS and R boundary conditions around all cycles. However, in an ungauged theory, there is a specific set of boundary conditions implied by the replica trick (just as the thermal partition function is computed with a specific boundary condition---namely antiperiodic---around the Euclidean time circle). We will not review the derivation here; it is given in \cite{Casini:2005rm} (below equation (6); see footnote 5 of \cite{Headrick:2010zt} for an alternate derivation), but the result is that for even $n$ one must include a sign-flip along the cuts that connect the first sheet to the last one. (These boundary conditions imply that, when passing to a single-valued coordinate system in the neighborhood of a branch cut, there is no spin field inserted at the branch point. They also imply NS boundary conditions on all the basis cycles we will use in Section 3, shown in figure \ref{canonical_cycles_fig}.) Since no sum over boundary conditions was performed, it is clear that the calculation in \cite{Casini:2005rm} is done in the ungauged theory. Indeed, in the case $N=n=2$, where the replicated surface is a torus, one can directly reproduce the CFH result from the well-known torus partition function for a fermion with NS boundary conditions on both cycles. We will do this in subsection 3.1 below, after reviewing the transformation from the flat torus to the singular surface $\Sigma_{2,2}$.

Alternatively, Refs.\ \cite{Casini:2009vk,Casini:2009sr} compute the reduced density matrix directly in terms of the two-point functions of Dirac fermions.  It is clear that this calculation is for the ungauged fermion.  The Hilbert space is factorized into left- and right-moving excitations, which is not possible for the $\Z_2$ gauged fermion dual to the boson; for example, the modular invariant partition function (which we will review below) does not factorize into contributions from left- and right-movers. In addition, the ``modular Hamiltonian", whose exponential forms the reduced density matrix (see equation (28) of \cite{Casini:2009vk}), consists of products of chiral fermion operators at different points.  In the gauged theory, however, a single chiral fermion is not in the spectrum of local operators, as it is odd under $\Z_2$. Furthermore, the entanglement entropies calculated directly in this approach match those calculated via the replica trick.

The free Dirac fermion is a consistent quantum theory, but it is not modular invariant, and so cannot be dual to the modular-invariant theory of the free boson at any radius. This explains the discrepancy between the R\'enyi entropies. However, there is a curious equality between the R\'enyi entropies for two intervals for the free Dirac fermion and the self-dual boson. We now turn to explaining this fact.


\section{R\'enyi entropies for arbitrary intervals}\label{sec:renyi_entropies}

The strange coincidence in the entanglement R\'enyi entropies for two intervals ($N=2$) and all $n$, between the free Dirac fermion and the self-dual boson, challenges the ability of the entanglement entropies to distinguish theories. Before making any sweeping claims, we should  compare the values of $S_{n}(A)$ for $N>2$ in the two theories. Agreement or disagreement of $S_n(A)$ between the two theories amounts to agreement or disagreement of the partition functions $Z_{n,N}$ on the singular Riemann surface $\Sigma_{n,N}$. We will find on quite general grounds that the theories agree for $n = 2$ and any $N$ (a new result), as well as for $N = 2$ and any $n$ \cite{Casini:2005rm,Casini:2009vk,Casini:2009sr}, for reasons which fail when $n \geq 3$ and $N\geq 3$. We check by direct numerical computation that the R\'enyi entropies differ when $n = N = 3$. Hence the full set of R\'enyi entropies does distinguish between the theories. Along the way we will discover a surprising relationship between the theories, which will explain in a simple way why certain R\'enyi entropies agree.

To compute the partition function, we can use a Weyl transformation to map the metric $ds^2$ on our singular surface $\Sigma_{n,N}$ to a non-singular fiducial metric $d\hat s^2 = e^{-\phi}ds^2$ \cite{Lunin:2000yv,Headrick:2010zt,Calabrese:2009qy,Calabrese:2009ez}. In this case,
\be
Z_{n,N} = e^{S_L} \hat Z_{n,N}\, .
\ee
Here $\hat Z$ is the partition function of the CFT with the fiducial metric, and $S_L$ is the Liouville action
\be\label{Liouville}
S_L = \frac{c}{96 \pi} \int \sqrt{\hat g} \left(\hat g^{ab} \partial_a\phi \partial_b \phi + 2 \hat R \phi \right).
\ee 
The Liouville action depends on the CFT solely via its central charge.  Since the Dirac fermion and free boson both have $c=1$, agreement of the R\'enyi entropies is equivalent to agreement of the partition functions on the non-singular Riemann surface. For the fermionic theory, it will also be important to keep track of boundary conditions around the various non-contractible cycles.

In Appendix \ref{app:genl_boson_renyi}, we apply the technology developed in this section to the compact boson at arbitrary radius, giving a relatively simple derivation of Calabrese, Cardy, and Tonni's result \eqref{CCT} for its R\'enyi entropies.

\subsection{Torus partition functions and momentum lattices}\label{subsec:torus}

The Riemann surface $\Sigma_{N,n}$ has genus $g=(N-1)(n-1)$. We begin with the simplest non-trivial case, $\Sigma_{2,2}$, which is a torus. The modular parameter $\tau$ of the torus depends on the cross-ratio $x$ of the endpoints of the intervals (the branch points in $\Sigma_{2,2}$), defined in \eqref{crossratio}, which lies in the range $0<x<1$ when both cuts are on the real line. The relation between $x$ and $\tau$ is
\begin{eqnarray}
\tau &=& \frac{iK(1-x)}{K(x)}\,,\qquad K(x) = \frac\pi2\,{}_2F_1\left(\frac{1}{2},\frac{1}{2},1;x \right),\nonumber\\
x &=& \frac{\vartheta_2^4(\tau)}{\vartheta_3^4(\tau)}\,,
\label{eq:torus_period}
\end{eqnarray}
where $K(x)$ is the complete hyperelliptic integral and ${}_2F_1$ is the usual hypergeometric function. The expressions \eqref{eq:torus_period} are actually valid for arbitrary complex $x$, but for $0<x<1$, $\tau$ is imaginary and the torus is rectangular. In this case, the partition function on a flat torus depends only on the spectrum of scaling dimensions $\Delta$ of the CFT:
\be\label{eq:hampf}
\hat Z_{2,2} = \tr e^{-2\pi \tau_2 H} = e^{2\pi\tau_2/12}\sum_me^{-2\pi\tau_2\Delta_m}
\ee
(where the sum is over operators $\mathcal{O}_m$ and $\tau_2=\Imag\tau$). The cycle playing the role of the spatial circle here is the one that, on $\Sigma_{2,2}$, encircles one of the cuts while staying on one sheet, while the one playing the role of the time circle encircles the two middle branch points, passing from one sheet to the other. Based on the boundary conditions explained in subsection 2.5, both cycles have periodic boundary conditions for the fermions on $\Sigma_{2,2}$, which corresponds to antiperiodic (NS) boundary conditions on the flat torus. Hence the trace is over NS-NS sector states and does not include a factor of $(-1)^F$.

From the agreement of $S_{2}(A)$ for general $x$ between the fermion and the self-dual boson, it follows that the two theories have the same partition function on any rectangular torus, and hence that they have identical spectra of scaling dimensions. In this subsection we will explain this agreement; in the rest of this section we will then use what we've learned to explain the agreement for other values of $n$ (with $N=2$), and find out to what extent it generalizes to other values of $(n,N)$.

Let us first recall the structure of the operators in the two theories. The Dirac fermion consists of separate left- and right-moving Weyl fermions $\psi_{L,R}$, which have conformal weights $h=1/2$ and $\tilde h=1/2$ respectively. The general operator is a product of distinct operators of the form $\partial^n\psi_L,\partial^n\bar\psi_L,\bar\partial^n\psi_R,\bar\partial^n\bar\psi_R$, where $n=0,1,\ldots$. The theory has conserved left- and right-moving fermion number currents.

The compact boson can be split into left- and right-moving bosons $X_{L,R}$. The exponential operators, which create winding and momentum ground states, are of the form $e^{ik_LX_L+ik_RX_R}$.\footnote{There are also cocycles, which we neglect here since they do not contribute to the scaling dimensions which are our main interest. In Section 4, where we will study correlators, we will include them.} The left- and right-moving momenta $k_{L,R}$ are not independent, but are elements of a joint lattice $\Gamma^{\rm (b)}$, which at the self-dual radius is as follows:
\begin{equation}\label{Gammabdef}
\Gamma^{\rm (b)} = \left\{(k_L,k_R):k_L\pm k_R\in\sqrt2\Z\right\}\,.
\end{equation}
The exponential operator has conformal weights $h=k_L^2/2$, $\tilde h=k_R^2/2$. The total momentum and the winding number are given in terms of $k_{L,R}$ by $n,w=(k_L\pm k_R)/\sqrt2$. The general operator is a product of an exponential  and derivative operators $\partial^nX_L,\bar\partial^nX_R$, $n=1,2,\ldots$. (The self-dual boson actually has a larger, $SU(2)\times SU(2)$ symmetry group, but we will have occasion only to use its momentum and winding $U(1)\times U(1)$ subgroup.)

For low-lying operators, it is straightforward to see by inspection that the spectra of the two theories are the same. For example, both theories have 4 dimension-1/2 operators (fermion: $\psi_{L,R}$, $\bar\psi_{L,R}$; boson: the $(n,w)=(\pm1,0),(0,\pm1)$ exponential operators) and 6 dimension-1 operators (fermion: $\psi_L\psi_R$, $\bar\psi_L\bar\psi_R$, $\psi_{L,R}\bar\psi_{L,R}$; boson: the $(n,w)=(\pm1,\pm1)$ exponentials and $\partial X_L,\bar\partial X_R$). Furthermore, it is possible to match not only the scaling dimensions but also the two theories' respective $U(1)\times U(1)$ charges, i.e.\ to establish a one-to-one correspondence such that the left- and right-moving fermion numbers match $n$ and $w$ respectively. On the other hand, the spins definitely cannot be made to match, given that one theory contains fermions and the other doesn't.

\begin{figure}[thbp]
\begin{center}
\includegraphics[scale=.6]{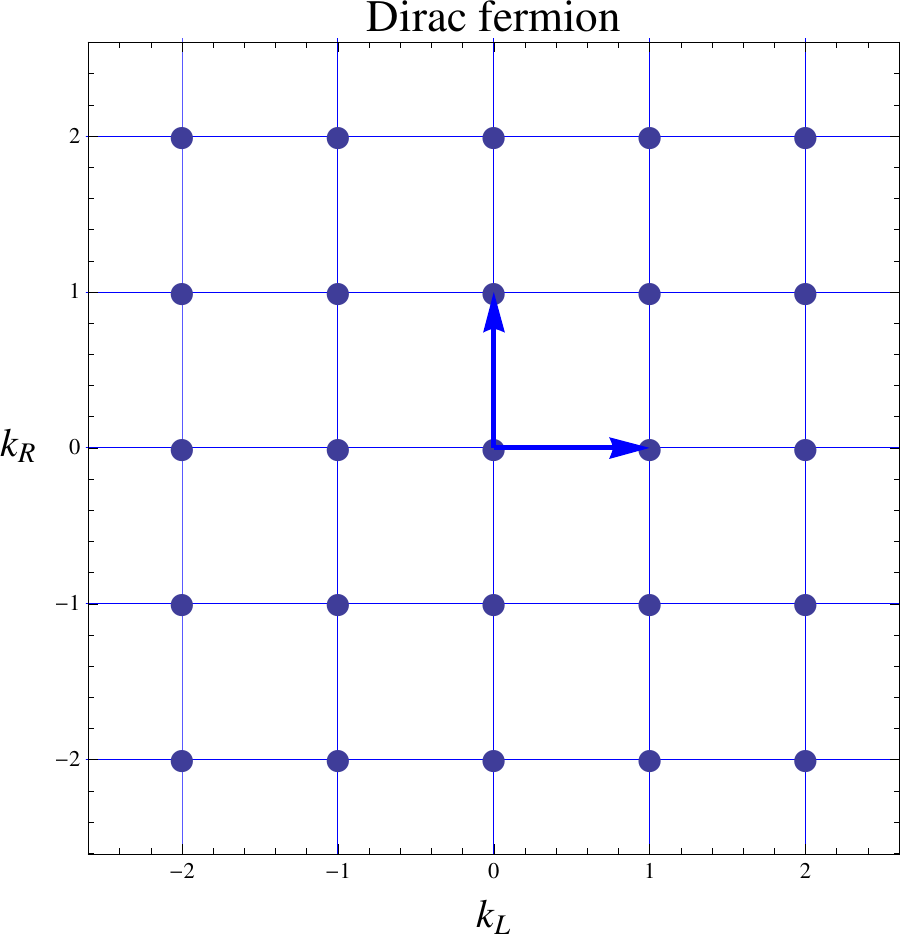}
\includegraphics[scale=.6]{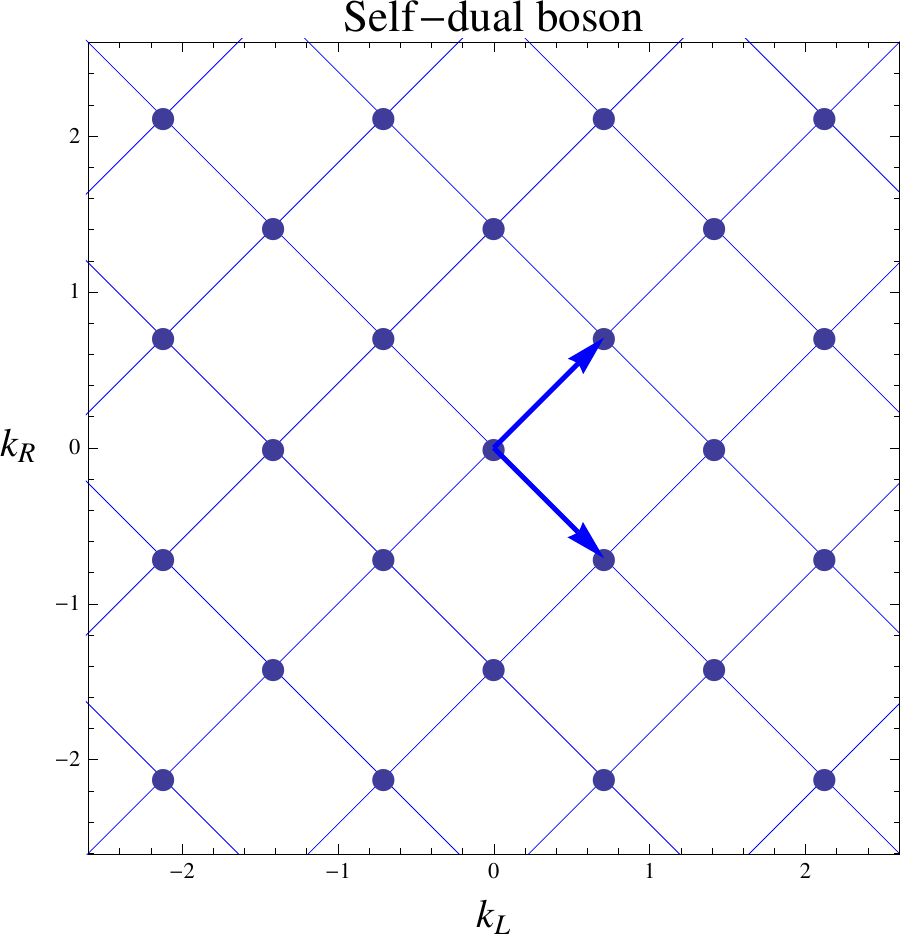}
\caption{The momentum lattices for the ungauged Dirac fermion (left) and for the self-dual boson (right). We have bolded the canonical generating vectors for both lattices. Notice that they are related by a 45$^\circ$ rotation.}
\label{momentum_lattice}
\end{center}
\end{figure}

It is not necessarily clear from these low-lying examples, however, what the general pattern is. The mystery is readily solved by bosonizing the fermion.\footnote{By "bosonizing" we mean representing the chiral fermion operators as exponentials of chiral boson operators.  The chiral bosons are defined via their OPEs. This is distinct from the bosonization of the modular-invariant fermion theory.}  In its bosonized form, the Dirac fermion consists of left- and right-moving bosons $H_{L,R}$, related to the elementary fermionic fields by
\be
\psi_L = e^{iH_L}\,,\qquad\psi_R = e^{iH_R}\,.
\ee
Just as for the self-dual boson, the general operator is written as a product of an exponential operator $e^{ik_LH_L+k_RH_R}$ and derivative operators $\partial^nH_L,\bar\partial^nH_R$. But in this case the momenta $k_{L,R}$, which are the left- and right-moving fermion numbers, are independent integers; in other words, the momentum lattice is simply
\begin{equation}
\Gamma^{\rm (f)} = \Z^2\,.
\end{equation}
(Note that we use the superscript (f) to refer to the Dirac fermion theory even when we are working with its bosonized form.) The two lattices are shown in figure \ref{momentum_lattice}, and it is immediately seen that they are related by a 45${}^\circ$ rotation. This rotation matches the $U(1)\times U(1)$ charges of the respective theories to each other:\footnote{We could just as well make other choices, like $(k_L^{\rm (f)},k_R^{\rm (f)}) = (n,-w)$ or $(w,n)$, but these are all related by automorphisms of the two theories, and therefore equivalent.}
\begin{equation}\label{45rotation1}
\left(k_L^{\rm (f)},k_R^{\rm (f)}\right) = 
\left(\frac{k_L^{\rm (b)}+k_R^{\rm (b)}}{\sqrt{2}},\frac{k_L^{\rm (b)}-k_R^{\rm (b)}}{\sqrt{2}}\right) = (n,w)\,.
\end{equation}
It also preserves the lengths of vectors defined with respect to the Euclidean inner product
\begin{equation}\label{eq:dot_defns}
k\cdot k' = k_Lk_L' + k_Rk'_R\,,
\end{equation}
and therefore the scaling dimensions $\Delta = k\cdot k/2$ of exponential operators; however it does not preserve the Lorentzian inner product
\begin{equation}
k\circ k' = k_Lk_L' - k_Rk'_R
\end{equation}
which gives their spins $s=k\circ k/2$.\footnote{The Lorentzian inner product is perhaps more familiar in the context of momentum lattices.  It is with respect to this inner product that, in string theory (for example in a Narain compactification), one requires the lattices to be integral (for mutual locality of operators), self-dual (for modular invariance under $\tau\to-1/\tau$), and even (for modular invariance under $\tau\to\tau+1$). $\Gamma^{\rm (f)}$ and $\Gamma^{\rm (b)}$ are both integral and self-dual, but only $\Gamma^{\rm (b)}$ is even.} In addition to having momentum lattices that are related by a rotation, the two theories have isomorphic sets of derivative operators: $\partial^nX_L,\bar\partial^nX_R$ for the bosonic theory and $\partial^nH_L,\bar\partial^nH_R$ for the fermionic theory all contribute $n$ to the scaling dimension and 0 to the $U(1)\times U(1)$ charge of an operator. Together these two facts explain the matching of the spectra of scaling dimensions as well as $U(1)\times U(1)$ charges.

Let us return to the torus partition function, which can be written in terms of a sum over the momentum lattice; this will be a useful warm-up for the higher-genus partition functions we will consider in the rest of this section. We will not assume that $\tau$ is imaginary. For the fermion we have
\be
\hat Z_{2,2}^{\rm (f)}(\tau, \bar\tau) = \frac{1}{|\eta(\tau)|^2} \sum_{k \in \Gamma^{\rm (f)}} \exp\left(i \pi \tau_1k\circ k - \pi\tau_2k\cdot k \right).\label{eq:dirac_torus}
\ee
The factor of $|\eta(\tau)|^{-2}$ accounts for the sum over all the possible derivative operators that can multiply a given exponential (i.e.\ the oscillators, in terms of states). For the boson we have almost the same formula:
\be
\hat Z_{2,2}^{\rm (b)}(\tau, \bar\tau) = \frac{1}{|\eta(\tau)|^2} \sum_{k \in \Gamma^{\rm (b)}} \exp\left(i \pi \tau_1k\circ k - \pi\tau_2k\cdot k \right).
\label{eq:general_boson_torus}
\ee
Again, since $\Gamma^{\rm (f)}$ and $\Gamma^{\rm (b)}$ are related by a transformation that preserves the Euclidean inner product \eqref{eq:dot_defns}, the two partition functions will agree precisely when $\tau_1=0$. In fact, this will work for any two lattices that are related by such an orthogonal transformation. However, if one restricts to integral self-dual lattices in two dimensions, then $\Gamma^{\rm (f)}$ and $\Gamma^{\rm (b)}$ are the only examples related in this way.

From \eqref{eq:dirac_torus} we can easily recover the CFH result \eqref{CasiniHuerta} for $N=n=2$, as promised in subsection 2.5. Taking $\tau$ imaginary, and using \eqref{eq:torus_period}, we have
\begin{equation}\label{diractorus2}
\hat Z_{2,2}^{\rm (f)}(\tau) = \frac{\vartheta_3(\tau)^2}{\eta(\tau)^2} = \left(\frac{2^4}{x(1-x)}\right)^{1/6}\,.
\end{equation}
The Liouville action \eqref{Liouville} for the Weyl transformation from the flat torus to the singular surface $\Sigma_{2,2}$ (which has conical singularities at the four branch points) was computed by Lunin and Mathur \cite{Lunin:2000yv}:
\begin{equation}\label{LuninMathur}
e^{S_L} = \left(\frac{(v_1-u_1)(v_2-u_2)}{\epsilon^2}\right)^{-1/4}\left(\frac{x^2}{2^8(1-x)}\right)^{1/12}\,.
\end{equation}
We obtain
\begin{equation}
Z_{2,2} = e^{S_L}\hat Z_{2,2} =
\left(\frac{(v_1-u_1)(v_2-u_2)(1-x)}{\epsilon^2}\right)^{-1/4}
\end{equation}
yielding \eqref{CasiniHuerta}.\footnote{We can also calculate the result for the modular-invariant gauged Dirac theory. Using the fact that, in that case, $\hat{Z} = \frac{\vartheta_3^2+\vartheta_2^2+\vartheta_4^2}{2\eta^2}$, we find $\F_2(x) = \frac{1+\sqrt{x}+\sqrt{1-x}}{2}$, which agrees with (\ref{CCT}) for $n=2,~\eta=2$, using the resummation identity $\vartheta_2^2(\tau)+\vartheta_3^2(\tau)+\vartheta_4^2(\tau)=2\vartheta_3(2\tau)\vartheta_3(\tau/2)$. }

In order to go to $n>2$ and/or $N>2$, we need to consider the partition functions of the theories on higher-genus Riemann surfaces, which we will do in subsection \ref{subsec:partition}. To have the necessary language, however, we first need to review some algebraic geometry.

\subsection{Some algebraic geometry background}\label{subsec:algebraic_geometry}

In order to set up the computation of the partition functions ${\hat Z}_{N,n}$, in this subsection we will describe the particular Riemann surfaces we are studying, and review some basic facts about Riemann surfaces that we will need.  More complete reviews of the relevant mathematics can be found in \cite{AlvarezGaume:1986es,AlvarezGaume:1987vm,farkas1992riemann}.

\begin{figure}[thbp]
\begin{center}
\includegraphics[scale=.55]{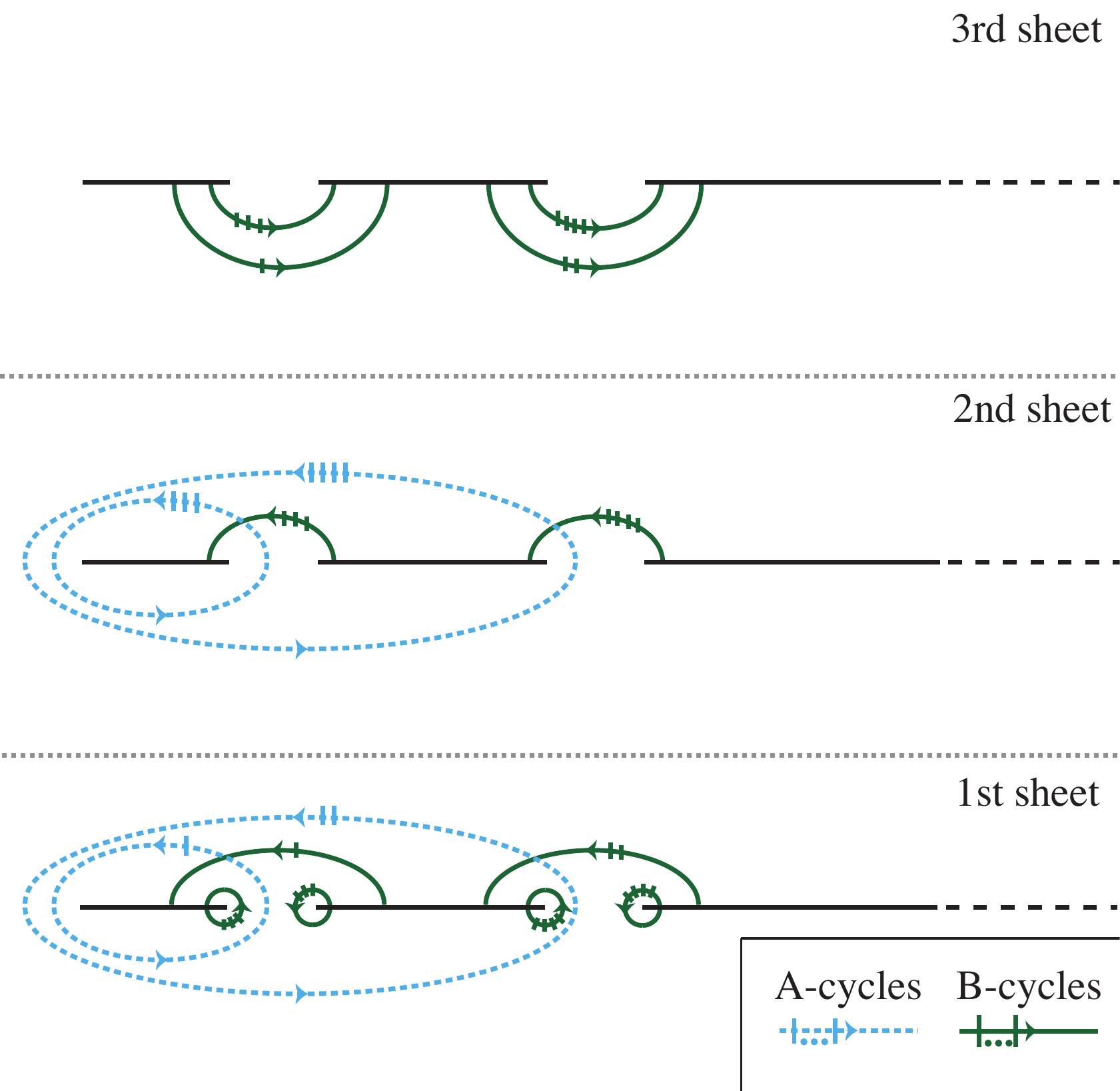}
\caption{The Riemann surface $\Sigma_{3,3}$ with a canonical basis of A- and B-cycles. The solid black lines are the branch cuts, oriented so that approaching from below takes one up a sheet and approaching from below takes one down a sheet. The blue dashed lines are A-cycles and the green solid lines are B-cycles, with notches corresponding to the index (eg $a_3,~a_4$ lie on the second sheet). We have pulled the B-cycles off of the branch points for clarity.}
\label{canonical_cycles_fig}
\end{center}
\end{figure}

The singular Riemann surfaces were described in subsection \ref{subsec:disc_gauge_sym}, and are illustrated in Fig. \ref{canonical_cycles_fig}\ for $n = N = 3$. The Riemann surface can be represented by the algebraic curve \cite{Enolski:2004a,Enolski:2006a} (see also Appendix C of \cite{Calabrese:2009ez}):
\be\label{eq:singularcurve}
	y^n = \prod_{k = 1}^{N-1} (z - u_k)(z - v_k)^{n-1}(z - u_N)
\ee
where $u_k,v_k$ all lie on the real line, and the conformal invariance has been used to send $v_N$ to $\infty$.  The $k$th branch cut lies between branch points at $u_k, v_k \in \RR$ with $u_k < v_k$.  The residual conformal invariance can be used to fix the location of two more of those points.  For instance, one can set the first interval to lie between $u_1 = 0$ and $v_1 = 1$.

The Riemann-Hurwitz formula gives the genus of this Riemann surface as $g = (N-1)(n-1)$.  On any Riemann surface, one can write down a canonical basis of ``A-cycles" $a_{i= 1 \ldots g}$ and ``B-cycles" $b_{i = 1,\ldots g}$ with intersection pairing $a_i \cdot b_j = \delta_{ij} = - b_j \cdot a_i$.   For the curve (\ref{eq:singularcurve}), the canonical basis we will use is as follows: labeling the cycles by  sheet $s = 1,\ldots, n-1$ and cut $k = 1,\ldots,N-1$, cycle $a_{k + (s-1)(N-1)}$ encircles the first $k$ cycles on sheet $s$ in a counterclockwise direction; cycle $b_{k + (s-1)(N-1)}$ passes from $v_k$ to $u_{k+1}$ on sheet $s$, and returns on sheet $N$ \cite{Enolski:2004a,Enolski:2006a}. This basis is shown in figure \ref{canonical_cycles_fig} for the case $n=N=3.$ Based on the boundary conditions for the Dirac fermion explained in subsection 2.5, all of these basis cycles carry NS boundary conditions, which we can see as follows. First, each one crosses an even number of sign flips, and hence carries periodic boundary conditions on this singular surface. Second, each one has odd winding number. Hence, under a Weyl transformation that takes the surface to a non-singular one, the fermion will be antiperiodic along the geodesic representative of each cycle.

There is  a canonical basis of holomorphic differentials $\zeta_i$ such that 
\be
	\oint_{a_i} \zeta_j = \delta_{ij}\ ;\ \ \ \ \ \oint_{b_i} \zeta_j = \Omega_{ij}
\ee
$\Omega_{ij}$ is known as the period matrix; it is symmetric, and the imaginary part of $\Omega_{ij}$ is a positive definite quadratic form.

A natural but noncanonical basis of holomorphic differentials is \cite{Enolski:2004a,Enolski:2006a}:
\be\label{eq:holbas}
\omega_{j+(N-1)s} = \frac{z^{j-1}\prod_{i=1}^{N-1}(z-v_i)^s}{y^{s+1}} dz, \qquad j = 1,\ldots, N-1,~s = 1,\ldots, n-1,
\ee
where $y$ is given in (\ref{eq:singularcurve}).  The period integrals are:
\be
A_{ij} = \oint_{a_i}\omega_j,~B_{ij}  = \oint_{b_i} \omega_j,~ \Omega = A^{-1} B.
\ee
The canonical basis is then $\zeta_j = \sum_{k} \omega_k A^{-1}_{kj}$.  As a warmup, one can use the basis (\ref{eq:holbas})\ to reproduce (\ref{eq:torus_period}) for genus one, where $\Omega_{11}=\tau$.

\subsection{Partition functions at higher genus}
\label{subsec:partition}

The partition function of the self-dual boson can be computed following \cite{AlvarezGaume:1986es,AlvarezGaume:1987vm,Dijkgraaf:1987vp}:
\be\label{eq:sdpfg}
	 \hat Z^{\rm (b)} = \pi\sqrt{\frac{2{\rm Vol}(\Sigma) \Imag\det \Omega}{\det' \nabla^2}}\sum_{\vec\eps \in \Z_2^g}
		\left| \vartheta\left[ \begin{array}{c}\vec \eps \\ 0 \end{array} \right](0 | 2 \Omega)\right|^2\,.
		\ee
Here ${\rm Vol}(\Sigma)$ is the volume of the surface; $\vec\epsilon\in\Z_2^g$ means $\epsilon^i \in \{0,1/2\}$;  $\det'\nabla^2$ is a suitably regularized determinant of the Laplacian; the factor of $\sqrt{2}\pi$ comes from the path integral over the zero mode of the boson; and  $\vartheta$ is the Riemann-Siegel theta function:
\be\label{eq:rsdef}
	\vartheta\left[\begin{array}{c} \vec\eps_1\\ \vec\eps_2 \end{array}\right] (\vec{z}|\Omega) 
	= \sum_{\vec{n} \in \Z^g}
	\exp\left[ i \pi ( n + \eps_1)^i \Omega_{ij} (  n + \eps_1)^j + 2\pi i ({ n} + { \eps}_1)^k({ z} + { \eps}_2)^k\right].
\ee

The Dirac fermion with antiperiodic (NS) boundary conditions about all basis cycles can also be computed following \cite{AlvarezGaume:1986es,AlvarezGaume:1987vm}.  Those works do the calculation in the $R = 1/2$ presentation of the dual boson, which can be written as a sum over spin structures of the fermion partition function. One can extract from this the partition function in the purely NS sector:
\be\label{eq:diracpfg}
	 \hat Z^{\rm (f)} = \pi\sqrt{\frac{2^g {\rm Vol}(\Sigma) \Imag \det \Omega}{\det' \nabla^2}} \left|\vartheta\left[\begin{array}{c} 0\\ 0 \end{array}\right] (0|\Omega)\right|^2
\ee
(which, again, includes the integral over the zero mode of the dual boson, equal to $\pi$).

To compare these partition functions, we first write $\Omega = A+iK,$, with $A,K$ symmetric and real and $K$ positive-definite. Using the results in Appendix A, we can rewrite (\ref{eq:sdpfg}) as a sum over $g$ copies of the momentum lattice $\Gamma^{\rm (b)}$ for the self-dual boson, $ k_L \pm k_R \in \sqrt{2}\Z$:
\be
\hat Z^{\rm (b)} =\pi\sqrt{\frac{2{\rm Vol}(\Sigma) \Imag\det \Omega}{\det' \nabla^2}}\sum_{\vec k \in (\Gamma^{\rm (b)})^g} \exp \left( \pi i A_{ij} k^i\circ k^j - \pi K_{ij} k^i \cdot k^j\right)\,.
\label{eq:sd_lattice_high_genus}
\ee
Similarly, the expression (\ref{eq:diracpfg}) can be written directly as a sum over $g$ copies of the momentum lattice $\Gamma^{\rm (f)} = \Z^2$,
\be
\hat Z^{\rm (f)} =\pi\sqrt{\frac{2^g{\rm Vol}(\Sigma) \Imag\det \Omega}{\det' \nabla^2}}\sum_{\vec k \in (\Gamma^{\rm (f)})^g} \exp \left( \pi i A_{ij} k^i\circ k^j - \pi K_{ij} k^i \cdot k^j\right)\,.
\label{eq:dirac_lattice_high_genus}
\ee
Since $\Gamma^{\rm (b)}$ is related to $\Gamma^{\rm (f)}$ by an orthogonal rotation, (\ref{eq:sd_lattice_high_genus}) and (\ref{eq:dirac_lattice_high_genus}) will agree (up to a factor of $2^{(g-1)/2}$) when $A_{ij}=0$---that is, when the period matrix is imaginary.  The factor of $2^{(g-1)/2}$ can be absorbed in a constant shift  of the dilaton coupling
\be
	\delta S_{\rm dilaton} = \frac{1}{4\pi} \int d^2 z \sqrt{g} R^{(2)} \Phi_0 = (2 - 2 g) \Phi_0
\ee
where $R^{(2)}$ is the worldsheet curvature. Furthermore, with the result as stated, this factor will only contribute to the non-universal part of the R\'enyi entropy as a constant term, and will cancel out of the mutual R\'enyi information entirely. 

The equality of the two partition functions for $\RealPart\Omega = 0$ is the central result of this subsection. We will now return to the Riemann surfaces $\Sigma_{n,N}$ involved in the calculation of R\'enyi entropies, and ask whether $\Omega$ is in fact imaginary for them.

\subsection{Symmetry constraints on the R\'enyi entropies}\label{subsec:dihedral_sym}

For the Riemann surfaces $\Sigma_{n,N}$, we will show that the real part of the period matrix vanishes identically when $N = 2$, and also when $n = 2$, but not otherwise.  Thus all of the R\'enyi entropies for two intervals will agree between the free Dirac fermion and the self-dual boson, consistent with the results in \cite{Casini:2009sr}. In addition, the second R\'enyi for any number of intervals will also agree; in other words, the CFH result \eqref{CasiniHuerta} for the Dirac fermion also applies to the self-dual boson for $n=2$, which we believe is a new result.

These facts arise because of the symmetries of the underlying Riemann surface.  The Riemann surface $\Sigma_{n,N}$ is a cyclic branched cover with all of the branch points on the real line.  Thus cyclic permutations of the sheets of the branched cover, as well as with complex conjugation, are symmetries of the surface. Together these symmetries form the dihedral group $D_n$. The period matrix transforms in a specific reducible representation $R_{\Omega}$ of $D_n$.\footnote{General treatments of real algebraic curves can be found in \cite{gross:1981a,Seppala:1989a}.  These use a different canonical homology basis than the one discussed here, related by a symplectic modular transformation.  Since the free Dirac fermion theory is not invariant under modular transformations, the partition function in that basis will not directly give the related R\'enyi entropy.}

These symmetries map A-cycles to A-cycles and B-cycles to B-cycles. Thus, they preserve the canonical structure of the period matrix. The upshot is that the period matrix should satisfy the conditions
\be
	g^T \Omega g = \Omega, ~ \tau \Omega \tau = - \Omega^*.
\ee
where $g$ is the $(N-1)(n-1)$-dimensional representation of the cyclic permutations acting on the $b$-cycles, and $\tau$ is the representation of the complex conjugation operators.  This implies that:
\be
	g^T K g = K,~\tau K \tau = K,~g^T A g = A,~\tau A \tau = - A.
\ee
In other words, every nonzero component of $A$ should lie in the one-dimensional irreducible representation of $D_n$ with $g$ acting trivially and $\tau = -1$; every nonzero component of $K$ should lie in the trivial representation of $D_n$. Our goal is to deduce the number of times these irreducible representations appear in $R_{\Omega}$.

To begin with we wish to deduce the representation $R_b$ of $D_n$ on the B-cycles, decomposed into irreducible representations of $D_n$, which we review in Appendix \ref{app:dihedral_rep}.  Consider first the case of 2 cuts.  The generator $g$ of the cyclic group $\Z_n$ simply permutes the sheets of the branch cover.  This, it acts on the B-cycles $b_{s = 1,\ldots,n-1}$ as
\be
	g = \left( \begin{array}{ccccc} -1 & -1 &  \ldots & -1 & -1 \\ 1& 0 & \ldots & 0 & 0 \\ 0 & 1 & \ldots & 0 & 0 \\ \ldots & \ldots & \ldots & \ldots & \ldots \\
	0 & 0 & \ldots & 1 & 0 \end{array} \right)
\ee
The complex conjugation $\tau$ takes the complex conjugate of the first sheet; for $n$ even, it also acts as complex conjugation on the $\left(\frac{n}{2} + 1\right)$st sheet.  Acting on the additional sheets, it exchanges the $k$th sheet with the complex conjugate of the $\left(n-k+2\right)$nd sheet.  We can thus write the action on the B-cycles as:
\be
	\tau = \left( \begin{array}{ccccccc} 
	1 & 0 & 0 & \ldots & 0&0 \\ 
	-1 & -1 & -1 & \ldots & -1 & -1  \\
	 0 & 0 & 0 & \ldots & 0 & 1 \\ 
	 0 & 0 & 0 & \ldots & 1 & 0 \\ 
	\ldots & \ldots & \ldots & \ldots & \ldots & \ldots \\ 
	0 & 0 & 1 & \ldots & 0 & 0 
	\end{array} \right)
\ee
The eigenvectors of $g$ with eigenvalue $\lambda$ have the form
\be
	v_{\lambda} = \left(\begin{array}{c} \lambda^{n-2}\\ \lambda^{n-3} \\ \ldots \\ \lambda \\ 1 \end{array}\right)
\ee
such that
\be
	\sum_{k = 1}^n \lambda^k = 0
\ee
For $n$ odd, the solutions are $\lambda_k = e^{2\pi i k/n}$.  Based on the representations listed in Appendix \ref{app:dihedral_rep}, we deduce:
\be
	R_\Omega = R_1 \oplus R_2 \oplus \ldots \oplus R_{\frac{n-1}{2}} \qquad \text{($n$ odd).}
\ee
For $n$ even, the solutions are $\lambda_k = e^{2\pi i k/n}$.  These include $\frac{n-2}{2}$ pairs of roots of unity and $\lambda_{n/2} = -1$.  For $\lambda = -1$, the corresponding eigenvector of $g$, $v_{\lambda_{-1}}$, is an eigenvector of $\tau$ with eigenvalue $1$.  Thus, in this case 
\be
	R_\Omega = {\bf S} \oplus R_1 \oplus R_2 \oplus \ldots \oplus R_{\frac{n-2}{2}} \qquad \text{($n$ even).}
\ee

In the case of $\tr\rho^n$ for two cuts, the period matrix transforms as the symmetrized product $(R_\Omega \otimes R_\Omega)_{sym}$.  To work this out, we need to decompose products of irreducible representations of $D_n$. This can be simply worked out from the specific matrix forms of $g,\tau$ given in Appendix \ref{app:dihedral_rep}:
\begin{eqnarray}
	R_j \otimes R_k & = & R_{j+k} \oplus R_{j-k} \ \ n \neq m \nonumber \\
	R_j \otimes R_j & = & R_{2j} \oplus {\bf 1} \oplus {\bf T} \nonumber\\
	R_j \otimes {\bf S} & = & R_{\frac{n - 2k}{2}} \nonumber\\
	{\bf S} \otimes {\bf S} & = & {\bf 1} \label{eq:repproducts}
\end{eqnarray}
The first two hold for $n$ even or odd.  The representation ${\bf T}$ only occurs in the second line.  However, it corresponds to the antisymmetrized tensor product of $R_j$ with itself, and does not appear in $(R_\Omega \otimes R_\Omega)_{sym}$. Therefore, the real component of the period matrix is zero in the two-cut case.  The R\'enyi entropies of the free Dirac fermion for two disjoint intervals are then identical to those for the self-dual boson, consistent with the results of \cite{Casini:2009sr}.

Next, let us consider the case of 3 or more cuts.  As discussed in subsection \ref{subsec:algebraic_geometry} and Fig.~\ref{canonical_cycles_fig}, the basis of B-cycles can be indexed as $b_{k + (s-1)(N-1)}$.  In this case $D_n$ acts as the representation $R_{\Omega}$ on the indices $s$, and the full representation is $\oplus_{k = 1}^{N-1}R_{\Omega}$.  The period matrix can be written in $(n-1)\times (n-1)$ blocks.  The $N-1$ diagonal blocks are each in the representation $(R_\Omega \otimes R_\Omega)_{sym}$, and so have no real parts.  The symmetry of $\Omega$ equates the $(N-1)(N-2)/2$ off-diagonal blocks below the diagonal to those above the diagonal.  Each of these blocks transforms in the {\it unsymmetrized}\ representation $R_{\Omega}\otimes R_{\Omega}$.  There are $(n-2)/2$ occurrences of the irrep ${\bf T}$ in each off-diagonal block when $n$ is even, and $(n-1)/2$ occurrences in each off-diagonal block when $n$ is odd.  Thus, the expected number of real components of the period matrix is:
\begin{eqnarray}
	\frac{(N-1)(N-2)(n-1)}{4} & {\rm components}  & {\rm for}\ n\ {\rm odd} \nonumber\\
	\frac{(N-1)(N-2)(n-2)}{4} & {\rm components}  & {\rm for}\ n\ {\rm even} \label{eq:realcomponents}
\end{eqnarray}
The second line implies a new result: for the second R\'enyi entropy $S_2(A)$, the CFH result for the fermion \eqref{CasiniHuerta} \cite{Casini:2005rm,Casini:2009vk,Casini:2009sr} applies also to the self-dual boson.\footnote{A very similar calculation gives us the number of imaginary components of the period matrix. These must lie in the trivial representation ${\bf 1}$ of $D_n$. This gives $N(N-1)(n-1)/2$ imaginary components for $n$ odd, and $N(N-1)n/4$ components for $n$ even.} 

Outside of the two families $N=2$, any $n$ and $n=2$, any $N$, symmetry constraints do not prevent a real part of the period matrix.  We have computed the period matrix and partition functions for the case $N=3$, $n=3$ using Mathematica, both by using the built-in function {\tt SiegelTheta} and by explicitly computing the theta functions, truncating the infinite sum (\ref{eq:rsdef}) at $n^i = 5$. The two methods agreed with each other to six decimal places.  For a variety of locations of the branch cuts of the Riemann surface, we found a single real component, as predicted in (\ref{eq:realcomponents}), which varied with the locations of the branch points, and we found that the partition functions of the free Dirac fermion and self-dual boson differed at the percent level.  As a check, if we set the real component of the period matrix to zero and inserted this into the partition functions (\ref{eq:sdpfg},\ref{eq:diracpfg}), we found that they agreed to six decimal places. We conclude that, in the end, the entanglement spectrum \emph{does} discriminate between different theories. A natural question to ask now is, what structures in the CFT are being singled out by the $n, N \geq 3$ R\'enyi entropies?  We will give a conjecture at the end of the next section.


\section{A real duality}\label{sec:real_duality}

We saw in the last section that the free Dirac fermion theory and the self-dual compact boson have identical spectra of scaling dimensions. We also saw that this coincidence can most easily be understood by bosonizing the fermion to obtain left- and right-moving bosons with a momentum lattice, $\Z^2$, that is related to that of the self-dual boson by a 45${}^\circ$ rotation
\begin{equation}\label{45rotation}
\left(k_L^{\rm (f)},k_R^{\rm (f)}\right) = 
\left(\frac{k_L^{\rm (b)}+k_R^{\rm (b)}}{\sqrt{2}},\frac{k_L^{\rm (b)}-k_R^{\rm (b)}}{\sqrt{2}}\right) = (n,w)\,,
\end{equation}
which preserves the scaling dimension $\Delta = (k_L^2+k_R^2)/2$, but not the spin $s = (k_L^2-k_R^2)/2$.

Unlike a true duality (such as T-duality), which acts separately on the left- and right-movers of a CFT, the 45${}^\circ$ rotation we are discussing mixes left- and right-movers. Nonetheless, it does give identical results for the partition functions on whole families of Riemann surfaces.  Friedan and Shenker \cite{Friedan:1986ua}\ have suggested that the partition function, understood as a section of a line bundle over the moduli space of Riemann surfaces, for all genera, should define the conformal field theory (via its factorization limits). In this spirit, it is worth asking whether the 45${}^\circ$ rotation (\ref{45rotation}) might preserve some of the structure of the two CFTs, such as correlators and OPEs. The purpose of this section is to show that a large amount of the structure---in fact, all correlators and OPEs on the real line---is preserved.

Consider for example the two-point function of elementary fermionic fields $\psi_L,\bar\psi_L = e^{\pm iH_L}$:
\begin{equation}\label{psipsi}
\bev{\psi_L(z_1)\bar\psi_L(z_2)} = \frac1{z_1-z_2}\,.
\end{equation}
These operators are mapped by \eqref{45rotation} to the operators of the self-dual boson with momenta $\pm1$ and no winding; these are spinless operators, whose two-point function is
\begin{equation}\label{mommom}
\bev{e^{i(X_L(z_1)+X_R(\zb_1))/\sqrt{2}}e^{-i(X_L(z_2)+X_R(\zb_2))/\sqrt{2}}} = \frac1{|z_1-z_2|}\,.
\end{equation}
While the correlators do not match for general positions, we see that they do match whenever $z_1-z_2$ is real and positive. Of course, two-point functions of primary operators are dictated by their scaling dimensions, so the nontrivial question is whether the agreement extends to non-primaries and to higher-point functions. In fact, as we will show, it does. The only restriction is that they must be placed on the real axis (or any common horizontal line) in the same order that they are written in the expectation value; that is, for the correlator $\ev{\mathcal{O}_1(z_1)\mathcal{O}_2(z_2)\cdots}$ we require the $z_i$ to be real and $z_1>z_2>\cdots$. We will prove this correspondence, which we call a ``real duality", in subsection 4.1. We will begin with the exponential operators, which are related by \eqref{45rotation}, then explain how to generalize the correspondence to other operators in such a way that OPEs on the real line are preserved; this immediately implies that all $n$-point functions on the real line are preserved.

In subsection 4.2, we will show that the component $T_{xx}=T+\tilde T$ of the stress tensor is preserved by this correspondence. Using the OPE, this implies that action of the mixed Virasoro generators $L_n+\tilde L_n$ are preserved, generalizing the matching of scaling dimensions. From this we learn what kinds of conformal transformations commute with the correspondence.

Finally, in subsection 4.3 we will use the Schottky construction, which describes an arbitrary Riemann surface as a quotient of the Riemann sphere by a discrete subgroup of $SL(2,\C)$, to explain how the real duality is related to the equality of partition functions between the two theories on Riemann surfaces with imaginary period matrices.

Throughout this section we will refer to the ungauged Dirac fermion theory as the ``fermion theory" (although we will mainly work in its bosonized form), and the theory of the self-dual boson as the ``boson theory". We think our meaning will be clear.

\subsection{Correlators}

In this subsection we will define a one-to-one correspondence between the operators of the fermion and the boson theories, and show that under this correspondence arbitrary correlators in the two theories agree, as long as the operators are placed on the real axis (in the same order that they are written in the expectation value). To prove this, we will show that the OPEs on the real axis agree; the statement about correlators follows since an arbitrary $n$-point function can be reduced to 1-point functions by repeated application of the OPE. Since the OPE proof is a bit formal, it is perhaps useful to see the real duality in action first. So we begin in 4.1.1 by proving it by explicit calculation for exponential operators. Then in 4.1.2 we explain how to generalize the one-to-one correspondence from exponential operators to general operators. Finally, we give the OPE proof in 4.1.3.

\subsubsection{Exponential operators}

In both the fermion and the boson theories, the exponential vertex operators include a cocycle, which is necessary because the pure exponential operators have the wrong statistics. For example, in the fermionic theory $e^{iH_L}$ and $e^{iH_R}$ commute even though they represent fermionic operators, and in the bosonic theory $e^{i(X_L(z_1)+X_R(\zb_1))/\sqrt{2}}$ and $e^{i(X_L(z_2)-X_R(\zb_2))/\sqrt{2}}$ anticommute even though they represent bosonic operators. In the fermion theory, the complete vertex operators are\footnote{With respect to the circle product $k\circ k'=k_Lk_L'-k_Rk_R'$, the lattice $\Z^2$ for the fermion theory is integral and self-dual but not even (i.e.\ the theory contains fermionic operators). The standard cocyle prescription, described in textbooks, applies only to even lattices, and as far as we know there is no simple general expression for the cocycles for non-even lattices (see \cite{Goddard:1983at} for a discussion of this point). That this particular theory admits a simple expression for the cocycle is presumably due to the fact that its lattice, while not even, is simply related to an even one.}
\begin{equation}\label{C1expop}
\mathcal{V}^{\rm (f)}_k = (-1)^{k_Rp_L}e^{ik_LH_L+ik_RH_R}\,,
\end{equation}
where $p_L$ is the left-moving momentum operator (the operator whose eigenvalue is $k_L$), while in the boson theory they are
\begin{equation}\label{C2expop}
\mathcal{V}^{\rm (b)}_k = (-1)^{(k_L-k_R)(p_L+p_R)/2}
e^{ik_LX_L+ik_RX_R}\,.
\end{equation}

We will now compute the correlators of these vertex operators, and prove the agreement claimed above. In the fermion theory we have
\begin{equation}\label{C1npoint}
\bev{\mathcal{V}^{\rm (f)}_{k^1}(z_1,\zb_1)\mathcal{V}^{\rm (f)}_{k^2}(z_2,\zb_2)\cdots} =
\begin{cases}
\prod_{i<j}(-1)^{k_L^ik_R^j}(z_i-z_j)^{k_L^ik_L^j}(\zb_i-\zb_j)^{k_R^ik_R^j}\,,\quad&\sum_ik^i = 0 \\
0\,,\quad &\text{otherwise}
\end{cases}\,.
\end{equation}
The sign factors $(-1)^{k_L^ik_R^j}$ come from moving the cocyles past the exponentials to the left until they hit the vacuum. The rest comes from the expectation value of the exponentials. In the boson theory the result is similar:
\begin{multline}\label{C2npoint}
\bev{\mathcal{V}^{\rm (b)}_{k^1}(z_1,\zb_1)\mathcal{V}^{\rm (b)}_{k^2}(z_2,\zb_2)\cdots}\\ =
\begin{cases}
\prod_{i<j}(-1)^{(k_L^i+k_R^i)(k_L^j-k_R^j)/2}(z_i-z_j)^{k_L^ik_L^j}(\zb_i-\zb_j)^{k_R^ik_R^j}\,,\quad&\sum_ik^i = 0 \\
0\,,\quad &\text{otherwise}
\end{cases}\,.
\end{multline}
The momentum conservation conditions are linear, and therefore preserved by the rotation \eqref{45rotation}. The sign from the cocycles is clearly preserved by \eqref{45rotation}. Finally, when the $z_i-z_j$ are all real and positive, the rest of the multiplicands collapse to $(z_i-z_j)^{k^i\cdot k^j}$, and the exponents are again equal under the rotation. So indeed \eqref{C1npoint} and \eqref{C2npoint} are equal under \eqref{45rotation}.

\subsubsection{General operators}

We now wish to extend the correspondence from exponential operators to general operators, which are products of the operators $\mathcal{V}_k$ and derivative operators $\partial^nH_L,\bar\partial^nH_R$ (for the fermion) or $\partial^nX_L,\bar\partial^nX_R$ (for the boson). Operators that include derivatives are degenerate for fixed values of the momenta and scaling dimension, so we cannot be guided by matching those quantum numbers alone. Our guiding principle for dealing with such operators will be the following. Since the correlators of exponential operators match only when they are placed on the real axis (or any other common horizontal line, but for simplicity we will take the real axis), we will work entirely on the real axis. Now, by the equations of motion for the fields, the holomorphic and antiholomorphic derivatives $\partial,\bar\partial$ can be replaced by the $x$-derivative $\partial_x = \partial + \bar\partial$:
\begin{equation}
\partial^nH_L = \partial_x^nH_L\,,\qquad
\bar\partial^nH_R = \partial_x^nH_R\,,
\end{equation}
and similarly for $X_L,X_R$. Having written all derivatives in terms of $\partial_x$, we simply apply the same rotation in field space that yielded the map between the exponential operators, namely
\begin{equation}\label{correspondence}
H_L \leftrightarrow \frac1{\sqrt2}(X_L+X_R)\,,\qquad
H_R \leftrightarrow \frac1{\sqrt2}(X_L-X_R)\,.
\end{equation}
Thus for example we have
\begin{align}\label{dercorrespondence}
\partial^nH_L=\partial_x^nH_L &\leftrightarrow \frac1{\sqrt2}\partial_x^n(X_L+X_R)=\frac1{\sqrt2}(\partial^nX_L+\bar\partial^nX_R)\,,\nonumber\\
\bar\partial^nH_R = \partial_x^nH_R &\leftrightarrow \frac1{\sqrt2}\partial^n_x(X_L-X_R) = \frac1{\sqrt2}(\partial^nX_L-\bar\partial^nX_R)\,.
\end{align}

It is useful to note that the rotation \eqref{45rotation} on the momentum lattice, together with the rotation \eqref{dercorrespondence} on the derivative operators, clearly preserves the Zamolodchikov metric on the space of operators. Since, by definition, the Zamolodchikov metric is the correlator $G_{mn}=\ev{\mathcal{A}'_m(\infty)\mathcal{A}_n(0)}$ (where $\mathcal{A}'_m$ is the operator $\mathcal{A}_m$ in the $z'=1/z$ frame), this implies that $n$-point functions where one of the operators is at infinity are also preserved by the real duality. We will also make use of the matching of the Zamolodchikov metric in subsection 4.3.

\subsubsection{OPEs}

We will now show that the OPE of arbitrary operators at real positions $z_1,z_2$, with $z_1>z_2$, is preserved, i.e.\ if $\mathcal{F}^{\rm (f)}\leftrightarrow\mathcal{F}^{\rm (b)},\mathcal{G}^{\rm (f)}\leftrightarrow\mathcal{G}^{(b)}$ then
\begin{equation}
\mathcal{F}^{\rm (f)}(z_1,\zb_1)\mathcal{G}^{\rm (f)}(z_2,\zb_2)\leftrightarrow
\mathcal{F}^{\rm (b)}(z_1,\zb_1)\mathcal{G}^{\rm (b)}(z_2,\zb_2)\,.
\end{equation}
As mentioned above, the agreement of arbitrary correlators on the real axis follows from the agreement of OPEs.

All of these composite operators are defined via normal-ordering; it will be useful to indicate this explicitly. Recall that in a free theory the OPE of operators $:\mathcal{F}(z_1,\zb_1):$, $:\mathcal{G}(z_2,\zb_2):$ is derived by adding to $:\mathcal{F}(z_1,\zb_1)\mathcal{G}(z_2,\zb_2):$ all possible cross-contractions between $\mathcal{F}$ and $\mathcal{G}$, and then Taylor-expanding with respect to $z_1-z_2$ and $\zb_1-\zb_2$ inside the normal-ordered product. We will first show that the cross-contractions match, then that the Taylor expansions match, when $z_1-z_2$ is real and positive

In the fermion theory, a cross-contraction consists of replacing an $H_L$ in $\mathcal{F}$ and an $H_L$ in $\mathcal{G}$ with the propagator $-\ln(z_1-z_2)$, or a pair of $H_R$s with $-\ln(\zb_1-\zb_2)$; similarly in the boson theory with $X_L$ and $X_R$. Let us first see how this works for derivative operators, then we will give the general proof. For example, the OPE of the left-moving fermion number current $\partial H_L$ with itself is
\begin{align}\label{example1}
:\partial H_L(z_1)::\partial H_L(z_2):\, 
&= {:\partial H_L(z_1)\partial H_L(z_2):} - \partial_1\partial_2\ln(z_1-z_2) \nonumber \\
&= {:\partial H_L(z_1)\partial H_L(z_2):} - \frac1{(z_1-z_2)^2}\,.
\end{align}
The corresponding operator in the boson theory is the momentum current $(\partial X_L+\bar\partial X_R)/\sqrt2$. Multiplying it by itself, we have four terms; two of them have mixed holomorphic and antiholomorphic parts and therefore no cross-contractions, while the other two are of the same form as \eqref{example1}:
\begin{multline}\label{example2}
\frac12{:\partial X_L(z_1)+\bar\partial X_R(\zb_1):}{:\partial X_L(z_2)+\bar\partial X_R(\zb_2):} \\
= \frac12{:\left(\partial X_L(z_1)+\bar\partial X_R(\zb_1)\right)\left(\partial X_L(z_2)+\bar\partial X_R(\zb_2)\right):}
-\frac1{2(z_1-z_2)^2} - \frac1{2(\zb_1-\zb_2)^2}\,.
\end{multline}
Comparing the right-hand sides of \eqref{example1} and \eqref{example2}, clearly the operator parts match under \eqref{dercorrespondence}, while their c-number parts are equal whenever the $z_i$ are real. More generally, a contraction of $\partial^{n_1}H_L(z_1)$ with $\partial^{n_2}H_L(z_2)$ gives
\begin{equation}
-\partial_1^{n_1}\partial_2^{n_2}\ln(z_1-z_2) = \frac{(-1)^{n_1}(n_1+n_2-1)!}{(z_1-z_2)^{n_1+n_2}}\,,
\end{equation}
while the contraction of the corresponding operators in the boson theory gives
\begin{equation}
-\frac12\partial_1^{n_1}\partial_2^{n_2}\ln(z_1-z_2)-\frac12\bar\partial_1^{n_1}\bar\partial_2^{n_2}\ln(\zb_1-\zb_2) = \frac{(-1)^{n_1}(n_1+n_2-1)!}{2(z_1-z_2)^{n_1+n_2}}+\frac{(-1)^{n_1}(n_1+n_2-1)!}{2(\zb_1-\zb_2)^{n_1+n_2}}\,.
\end{equation}
Again, these are equal when the $z_i$ are real. Clearly the same thing will hold for the contraction of $\bar\partial^{n_1}H_R(\zb_1)$ with $\bar\partial^{n_2}H_R(\zb_2)$.

For the proof that the OPEs of general operators match, we now apply the general formula for the cross-contractions in a free field theory (see for example equation (2.2.10) in \cite{Polchinski:1998rq}). In the fermion theory this is
\begin{multline}
:\mathcal{F}(z_1,\zb_1)::\mathcal{G}(z_2,\zb_2): \\
= \exp\left[-\int d^2z_1d^2z_2\left(\ln(z_1-z_2)\frac\delta{\delta H_L(z_1)}\frac\delta{\delta H_L(z_2)}+\ln(\zb_1-\zb_2)\frac\delta{\delta H_R(\zb_1)}\frac\delta{\delta H_R(\zb_2)}\right)\right]\\ \times:\mathcal{F}(z_1,\zb_1)\mathcal{G}(z_2,\zb_2):\,,
\end{multline}
Since the operators involved are contained entirely on the real axis (including all derivatives, when written using $\partial_x$), we can replace the integrals and functional derivatives with respect to $z,\zb$ with ones with respect to $x$:
\begin{multline}
:\mathcal{F}(x_1)::\mathcal{G}(x_2):\,= \\
\exp\left[-\int dx_1dx_2\ln(x_1-x_2)\left(\frac\delta{\delta H_L(x_1)}\frac\delta{\delta H_L(x_2)}+\frac\delta{\delta H_R(x_1)}\frac\delta{\delta H_R(x_2)}\right)\right]:\mathcal{F}(x_1)\mathcal{G}(x_2):\,.
\end{multline}
Similarly, in the boson theory we have
\begin{multline}
:\mathcal{F}(x_1)::\mathcal{G}(x_2):\,= \\
\exp\left[-\int dx_1dx_2\ln(x_1-x_2)\left(\frac\delta{\delta X_L(x_1)}\frac\delta{\delta X_L(x_2)}+\frac\delta{\delta X_R(x_1)}\frac\delta{\delta X_R(x_2)}\right)\right]:\mathcal{F}(x_1)\mathcal{G}(x_2):\,.
\end{multline}
These clearly map to each other under \eqref{correspondence}. Note that we also need $x_1>x_2$, otherwise the branch cut in the logarithm can lead to a mismatch (as between \eqref{psipsi} and \eqref{mommom}). Finally, we note that moving the cocycle for $\mathcal{G}$ through $\mathcal{F}$ gives a factor of $(-1)^{k_L^\mathcal{F}k_R^{\mathcal{G}}}$ in the fermion theory and of $(-1)^{n^\mathcal{F}w^{\mathcal{G}}}$ in the boson theory, but these are equal under the correspondence.

Finally, continuing to work on the real axis and to express derivatives using $\partial_x$, it is clear that Taylor-expanding the normal-ordered operators obtained from the cross-contractions with respect to $x_1-x_2$ will commute with the map \eqref{correspondence}. Hence the correspondence we have described preserves the full OPEs of arbitrary operators on the real axis.

\subsection{Stress tensor and real conformal transformations}

The components of the stress tensor in the two theories are
\begin{equation}
T^{\rm (f)} = -\frac12\partial H_L\partial H_L\,,\qquad \tilde T^{\rm (f)} = -\frac12\bar\partial H_R\bar\partial H_R
\end{equation}
and
\begin{equation}
T^{\rm (b)} = -\frac12\partial X_L\partial X_L\,,\qquad \tilde T^{\rm (b)} = -\frac12\bar\partial X_R\bar\partial X_R\,,
\end{equation}
respectively. While these components do not individually map to each under the correspondence \eqref{dercorrespondence}, their sum (which is the component $T_{xx}$) does:
\begin{equation}
T^{\rm (f)}+\tilde T^{\rm (f)} \leftrightarrow T^{\rm (b)}+\tilde T^{\rm (b)}\,.
\end{equation}
The OPEs of $T,\tilde T$ with other operators determine the action of the Virasoro generators:
\begin{equation}
T(z)\mathcal{A}(0) = \sum_nz^{-(n+2)}L_n\cdot\mathcal{A}(0)\,,\qquad
\tilde T(\zb)\mathcal{A}(0) = \sum_n\zb^{-(n+2)}\tilde L_n\cdot\mathcal{A}(0)\,,
\end{equation}
where $L_n\cdot\mathcal{A}$ denotes the result of $L_n$ acting on the operator $\mathcal{A}$ via the state-operator mapping (i.e.\ if the state-operator mapping maps $\ket{\mathcal{A}}$ to $\mathcal{A}$, then it maps $L_n\ket{\mathcal{A}}$ to $L_n\cdot\mathcal{A}$). Since the OPEs between operators on the real axis are preserved by the correspondence, for corresponding operators $\mathcal{A}^{\rm (f)},\mathcal{A}^{\rm (b)}$, we have
\begin{align}
\sum_nx^{-(n+2)}\left(L_n^{\rm (f)}+\tilde L_n^{\rm (f)}\right)\cdot\mathcal{A}^{\rm (f)}(0) &= \left(T^{\rm (f)}(x)+\tilde T^{\rm (f)}(x)\right)\mathcal{A}^{\rm (f)}(0) \nonumber\\
&\leftrightarrow \left(T^{\rm (b)}(x)+\tilde T^{\rm (b)}(x)\right)\mathcal{A}^{\rm (b)}(0) \nonumber\\
&= \sum_nx^{-(n+2)}\left(L_n^{\rm (b)}+\tilde L_n^{\rm (b)}\right)\cdot\mathcal{A}^{\rm (b)}(0)\,.
\end{align}
Hence the action of $L_n+\tilde L_n$ commutes with the correspondence. This statement generalizes the fact that the scaling dimension, which is the eigenvalue of $L_0+\tilde L_0$, is preserved.

The actions of $L_n,\tilde L_n$ in turn determine how an arbitrary operator transforms under conformal transformations. Specifically, under an infinitesimal conformal transformation $z\mapsto z+\epsilon v(z)$, we have:
\begin{equation}
\delta\mathcal{A}(z,\zb) = -\epsilon\sum_{n=0}^\infty\frac1{n!}\left(v^{(n)}(z)L_{n-1} + v^{(n)}(z)^*\tilde L_{n-1}\right)\cdot\mathcal A(z,\zb)\,,
\end{equation}
where $v^{(n)}=\partial^nv$. Since the action of $L_n+\tilde L_n$ is preserved by the correspondence, if $v^{(n)}(z)$ is real for all $n$ (at the location of the operator) then
\begin{equation}
\delta\mathcal{A}^{\rm (f)} \leftrightarrow \delta\mathcal{A}^{\rm (b)}\,.
\end{equation}
Exponentiating an infinitesimal transformation such that $v^{(n)}(x)$ is real for all real $x$ yields a finite transformation described by a real analytic function $x'(x)$ with positive first derivative. Such ``real conformal transformations" are compatible with the correspondence between the two theories, in the sense that if $\mathcal{A}^{\rm (f)}(x)\leftrightarrow\mathcal{A}^{\rm (b)}(x)$ then $\mathcal{A}^{\prime\rm (f)}(x')\leftrightarrow\mathcal{A}^{\prime\rm (b)}(x')$. Note that real conformal transformations also preserve the order of the positions of operators along the real axis.

Just as the notion of a usual (complex) conformal transformation is local and can be used to do a coordinate transformation on a patch of a manifold, the same holds for real conformal transformations. A simple example is afforded by the cylinder defined by identifying the plane in the imaginary direction $w\sim w+2\pi i$. The conformal transformation $z=e^w$ maps the cylinder to the plane with the origin removed. The map induces a real conformal transformation from the real axis in the $w$-plane to the positive real axis in the $z$-plane. Therefore, since the correspondence between the fermion and boson theories preserves correlators on the real axis of the $z$-plane, it also preserves correlators on the real axis of the $w$-cylinder. Such correlators represent, in the Lorentzian theory, equal-time correlators at finite temperature.

\subsection{Partition functions and the real duality}

In Section \ref{sec:renyi_entropies} we showed that the partition functions of the Dirac fermion and the self-dual boson agree on Riemann surfaces with imaginary period matrices. In this section we have shown so far that, under a certain one-to-one mapping between the operators of the theories, the correlators on the real axis agree. Both results are essentially due to the fact that the momentum lattices for the two theories are related by a rotation that preserves the dot product, $k\cdot k'=k_Lk_L'+k_Rk_R'$. In this section we will argue that these results can also be directly related to each other, since a genus-$g$ partition function can be written in terms of $2g$-point functions on the plane. As we will review, there is a relationship between the period matrix and the positions of the operators which is such that the period matrix is imaginary if the positions are real. We conjecture that the reality of the positions is both sufficient and \emph{necessary} for the period matrix to be imaginary. If this conjecture is true, the agreement between the partition functions follows from the real duality.

We first briefly review the relation between the genus-$g$ partition function of a general CFT and $2g$-point functions on the plane, closely following the discussion in the appendices of \cite{Gaberdiel:2010jf}. We begin with the Schottky construction, which describes an arbitrary genus-$g$ Riemann surface as a quotient of the Riemann sphere by a discrete subgroup $\Gamma$ of $SL(2,\C)$. This is a free group with generators $\gamma_i$, $i=1,\ldots,g$, that act as follows:
\begin{equation}
\frac{\gamma_i(z)-a_i}{\gamma_i(z)-r_i} = p_i\frac{z-a_i}{z-r_i}\,;
\end{equation}
we have parametrized $\gamma_i$ in terms of its attractive and repulsive fixed points $a_i,r_i$ and the dilatation parameter $p_i$, which satisfies $0<|p_i|<1$. To obtain a fundamental domain for $\Gamma$, we remove from the plane, for each $i$, the discs
\begin{equation}
D_i = \left\{\left|\frac{z-a_i}{z-r_i}\right|<R_i\right\},\qquad
D_{-i} = \left\{\left|\frac{z-r_i}{z-a_i}\right|<R_{-i}\right\},
\end{equation}
where $R_{i,-i}$ are chosen so that
\begin{equation}
R_iR_{-i}=|p_i|\,,
\end{equation}
and so that none of the discs overlap. Consistency of these requirements places some restrictions on the Schottky parameters $a_i,r_i,p_i$. To reconstruct the Riemann surface, one identifies the boundaries of $D_{i,-i}$ by the action of $\gamma_i$. There is a fundamental basis of A and B-cycles, in which the A-cycles are represented by the $\partial D_i$, and the B-cycles by lines connecting $D_i$ to $D_{-i}$. With the restrictions mentioned above, the Schottky parameters cover the moduli space of genus-$g$ Riemann surfaces (with some redundancies; for example, one can conjugate the entire Schottky group by an element of $SL(2,\C)$ without changing the Riemann surface).

By the standard sewing construction, the partition function of a CFT on the surface obtained from the Schottky construction can be written as a sum of $2g$-point functions on the plane. We will simply quote the result here; the detailed derivation can be found in appendix C of  \cite{Gaberdiel:2010jf}. Let $\{\mathcal{A}_m\}$ be a basis of operators with conformal weights $h_m,\tilde h_m$ respectively, and let $G_{mn}$ be the Zamolodchikov metric and $G^{mn}$ its inverse. Then
\begin{multline}
\label{eq:pftocorr}
Z_g(p_i; a_i; r_i) = \sum_{m_1,n_1,\ldots,m_g,n_g}
\left(\prod_i p_i^{h_{m_i}} {\bar p}_i^{\tilde h_{m_i}}G^{m_in_i}\right)
\\ \times\bev{\prod_i
\left((r_i-a_i)^{L_0}(\bar r_i-\bar a_i)^{\tilde L_0}e^{L_1+\tilde L_1}\mathcal{A}_{m_i}(r_i)\right)
\left((r_i-a_i)^{L_0}(\bar r_i-\bar a_i)^{\tilde L_0}e^{-L_1-\tilde L_1}\mathcal{A}_{n_i}(a_i)\right)
}.
\end{multline}

We will now show that \eqref{eq:pftocorr} gives the same result for the partition function for the fermion and boson theories, whenever the $p_i,a_i,r_i$ are all real and the $p_i$ are all positive. First, in that case the formula simplifies as follows:
\begin{multline}
\label{eq:pftocorrreal}
Z_g(p_i; a_i; r_i) = 
\sum_{m_1,n_1,\ldots,m_g,n_g}
\left(\prod_i p_i^{\Delta_{m_i}}G^{m_in_i}(\sgn(r_i-a_i))^{2s_{m_i}}\right)
\\ \times\bev{\prod_i
\left(|r_i-a_i|^{L_0+\tilde L_0}e^{L_1+\tilde L_1}\mathcal{A}_{m_i}(r_i)\right)
\left(|r_i-a_i|^{L_0+\tilde L_0}e^{-L_1-\tilde L_1}\mathcal{A}_{n_i}(a_i)\right)
}.
\end{multline}
The factor of $(\sgn(r_i-a_i))^{2s_{m_i}}$ arises from writing $(r_i-a_i)^{L_0}(\bar r_i-\bar a_i)^{\tilde L_0}=|r_i-a_i|^{L_0+\tilde L_0}\sgn(r_i-a_i)^{L_0-\tilde L_0}$; since $\mathcal{A}_{m_i}$ and $\mathcal{A}_{n_i}$ necessarily have the same spin (otherwise their Zamolodchikov inner product would be zero), and $e^{\pm(L_1+\tilde L_1)}$ doesn't change their statistics, we can write the sign factor as $(\sgn(r_i-a_i))^{2s_{m_i}}$ and pull it out of the expectation value. From the real duality---specifically, the agreement of scaling dimensions, the Zamolodchikov metric, the action of $L_n+\tilde L_n$, and correlators on the real line---we see that \eqref{eq:pftocorrreal} gives the same result when applied to the fermion and boson theories. The factor of $(\sgn(r_i-a_i))^{2s_{m_i}}$ takes care of the fact that the $\mathcal{A}_{m_i}(r_i)$ and $\mathcal{A}_{n_i}(a_i)$ are in the ``wrong" order (for the real duality) when $a_i>r_i$; finally, the order of the multiplicands for different values of $i$ inside the expectation value doesn't matter, since $\mathcal{A}_{m_i}$ and $\mathcal{A}_{n_i}$ are either both bosonic or both fermionic, so the full multiplicand is always bosonic. Since they have the same central charge and therefore the same Weyl anomaly, they will also have the same partition function on any surface related to this one by a Weyl transformation, including the constant-curvature one.

We have now proven that the partition functions of the fermion and the boson theories are the same under two separate sets of conditions on the moduli of the Riemann surface:
\begin{enumerate}
\item when the period matrix $\Omega$ is imaginary (in Section \ref{sec:renyi_entropies});
\item when the Schottky parameters $a_i,r_i,p_i$ are all real and the $p_i$ are positive (just above).
\end{enumerate}
While we are not aware of a proof in the mathematical literature, it seems very likely that these two sets of conditions are actually equivalent, i.e.\ that the period matrix is imaginary precisely under the conditions (2) on the Schottky parameters (or rather, when the $p_i$ are positive and the $a_i,r_i$ can be chosen to be real using the $SL(2,\C)$ freedom).\footnote{We would like to thank M. Gaberdiel, R. Volpato, and X. Yin for helpful discussion on this point.} The relation between the Schottky parameters and the period matrix (in the basis of A- and B-cycles described above) is known explicitly, but is somewhat complicated:
\begin{eqnarray}
	e^{2\pi i \Omega_{ii}} & = & p_i \prod_{\gamma \in \langle \gamma_i\rangle \setminus \Gamma / \langle \gamma_i\rangle} \frac{(a_i - \gamma(a_i))(r_i - \gamma(r_i))}{(a_i - \gamma(r_i))(r_i - \gamma(a_i))} \nonumber\\
	e^{2\pi i \Omega_{ij}} & = & \prod_{\gamma \in \langle \gamma_i\rangle \setminus \Gamma / \langle \gamma_j\rangle} \frac{(a_i - \gamma(a_j))(r_i - \gamma(r_j))}{(a_i - \gamma(r_j))(r_i - \gamma(a_j))}\qquad(i\neq j)\,.\label{eq:modmap}
\end{eqnarray}
Here $\langle \gamma_i\rangle \setminus \Gamma / \langle \gamma_j\rangle$ is the set of all $\gamma \in \Gamma$, written as words made from the letters $\gamma_{k=i,\ldots g}$ and their inverses, such that the first letter is not $\gamma_i$ or $\gamma_i^{-1}$ and the last letter is not $\gamma_j$ or $\gamma_j^{-1}$. From \eqref{eq:modmap} we can almost prove one direction of the equivalence. If the Schottky parameters are real, then obviously $e^{2\pi i\Omega_{ij}}$ is real for all $i,j$. If in addition $p_i>0$, then $e^{2\pi i\Omega_{ii}}>0$, hence $\Omega_{ii}$ is imaginary (the product over $\gamma$ is a perfect square, hence positive, since the multiplicand has the same value for $\gamma$ and $\gamma^{-1}$). Presumably the same can be shown for the off-diagonal components of $\Omega$. Showing the converse, that if $\Omega$ is imaginary then the $p_i$ are positive and the $a_i,r_i$ can be chosen to be real, seems more challenging, and we will not attempt it here.


\section{Discussion}\label{sec:discussion}

\subsection{Generalizations}

\subsubsection{Other configurations and states}

So far in this paper we have taken the field theories being studied to be in their ground states. However, it is straightforward to generalize the analysis to finite-temperature states. In this case the Euclidean spacetime that gets replicated in the replica trick is periodically identified in the Euclidean time direction (with NS boundary conditions for fermions), giving a cylinder. Including the points at spatial infinity, this is a sphere, so the replicated surface has the same topology as at zero temperature. Although it has a different complex structure from the zero-temperature case, the same basis of cycles can be used, and the dihedral representation theory argument given in subsection \ref{subsec:dihedral_sym} goes through as before. Hence, just as at zero temperature, this Riemann surface has an imaginary period matrix for $N=2$ and for $n=2$, and therefore the R\'enyi entropies for the Dirac fermion and the self-dual boson agree in these cases.

Another generalization is to quantize the theories on a circle, rather than a line. In this case, the replicated surface is periodic in the spatial direction. At zero temperature we again have a cylinder, and if we put NS boundary conditions on the fermion then the analysis of the previous paragraph shows that we will again get agreement for $N=2$ and for $n=2$. On the other hand, if we consider the theories on a circle at finite temperature, then the Euclidean spacetime is a torus. The $n$-sheeted replicated surface now has an extra $2n$ cycles, which transform in the fundamental representation of the dihedral group. The representation theory is therefore the same as if we had added another cut (plus one extra trivial representation). Hence the R\'enyis will agree for $N=1$ and for $n=2$.

Returning to the theory on the line at zero temperature, a different generalization is to intervals in spacetime that do not lie on a constant-time line. These are more difficult to compute, since the usual replica trick cannot be applied. However, it is possible that these quantities are related, perhaps by some sort of analytic continuation, to Euclidean partition functions where the endpoints $u_i,v_i$ of the intervals are moved off the real axis. Since our explanation of the agreement between the R\'enyis for the Dirac fermion and the self-dual boson crucially required those branch points to be real (in particular in our analysis of the symmetries of the relevant Riemann surface in subsection \ref{subsec:dihedral_sym}), it seems very likely that the non-equal-time R\'enyis will indeed distinguish between the theories, even for two intervals.

\subsubsection{Entanglement negativity}
In recent work \cite{Calabrese:2012ew,Calabrese:2012nk}, the authors considered the computation of the R\'enyi entropies for the partial transpose of the reduced density matrix for two intervals.\footnote{We thank an anonymous referee for pointing out this work and the issues it raises relative to ours.} This involved computing the partition function on a Riemann surface which is similar to $\Sigma_{n,2}$ but with the sheets attached in the opposite order on one cut relative to the other. The cuts are still on the real axis, and the surface still has a dihedral symmetry, yet the period matrices computed in \cite{Calabrese:2012nk}\ are not purely imaginary. The reason that our argument in subsection \ref{subsec:dihedral_sym} fails in this case is that the B-cycles must be chosen differently on this surface, and the action of the dihedral group (in particular the antiholomorphic involution) mixes them with the A-cycles, whereas we have assumed that the dihedral group acts separately on the A- and B-cycles, as it does on $\Sigma_{N,n}$.

A related point concerns non-modular-invariant theories such as the Dirac fermion. The authors of \cite{Calabrese:2012nk}, which studied only modular-invariant theories, obtained the R\'enyis for the partial transpose by analytically continuing the partition function in the cross-ratio $x$ to real values outside the interval $0<x<1$. However, for non-modular-invariant theories, one must be more careful, as can be seen by considering the case $n=2$. For $n = N = 2$, the modular parameter $\tau$ obtained by this analytic continuation has a real part (specifically, $\tau_1=1$).  However, in any quantum system, the trace of the square of the density matrix and the trace of the square of its partial transpose are identical. Since the partition functions of the free Dirac fermion and self-dual boson do not generally agree when $\tau_1\neq0$, but must agree in this case, one may worry that there is a contradiction.  There is not: rather, the analytic continuation argument in \cite{Calabrese:2012nk} cannot be applied to the non-modular-invariant Dirac fermion.  The continuation gives NS boundary conditions around the cycles $a, b-a$, while a direct path integral argument shows that the correct boundary conditions are NS around the cycles $a,b$, even for the trace of the square of the partial transpose.

\subsubsection{Other pairs of theories}

The free Dirac fermion and the compact boson are among the simplest quantum field theories one can study. An obvious question is whether the coincidences we have found---concerning entanglement entropies, partition functions, and correlation functions---can occur for more complicated theories, or whether they are in some sense artifacts of these theories' simplicity.

One generalization to a class of more complicated theories follows straightforwardly from our analysis. The key relationship between the Dirac fermion and self-dual boson theories, which allowed us to show that their partition functions agreed for imaginary period matrices (hence their R\'enyi entropies for $N=2$ and for $n=2$), as well as to prove the real duality, was the fact that their momentum lattices $\Gamma^{\rm (f)},\Gamma^{\rm (b)}$ are related by a transformation that preserves the Euclidean inner product $k\cdot k'=k_Lk_L'+k_Rk_R'$. It is clear from our analysis that any two theories that can be described in terms of left- and right-moving bosons on momentum lattices related by an orthogonal transformation will enjoy the same set of coincidences. Trivial examples include T-duality (which takes $k_R\to -k_R$) and parity (which exchanges $k_L$ and $k_R$), but more interesting examples will mix left- and the right-movers. For integral self-dual two-dimensional lattices, $\Gamma^{\rm (f)}$ and $\Gamma^{\rm (b)}$ furnish the only such example, but presumably with more bosons there are more examples. In fact, it would be interesting to see whether there are pairs of \emph{even} self-dual lattices related in this way. One could also consider lattices that are not integral or not self-dual.

More generally, it would be interesting to study whether similar ``partial" dualities can occur for theories that are not described in terms of free bosons, or in higher-dimensional theories.

The special role of dynamics on a codimension-one surface has the flavor of boundary conformal field theory \cite{Cardy:2004hm}, of which a standard  example is taking a CFT on the upper half plane with some boundary conditions on the real line. An exactly solvable BCFT in terms of an interacting self-dual scalar was described in \cite{Callan:1994ub}, and was nontrivially fermionized in \cite{Polchinski:1994my} (leading to a different theory than the non-modular invariant Dirac fermion). It would be very interesting to know if this nontrivial exact equivalence of theories is related in some way to our real duality, which is of course not an exact equivalence.

\subsection{Connections to larger issues}

Taking account of what we have learned in this paper, we return in this final subsection to the questions we posed at the beginning of the paper: Are entanglement entropies in quantum field theories (or, more precisely, their finite parts) universal quantities, and do they distinguish between theories?

\subsubsection{Position- vs.\ momentum-space entanglement}

The results of section \ref{sec:ent_entropy} are consistent with the universality of the entanglement spectrum of the reduced density matrix for spatial subsets of a field theory.  More precisely, the \emph{cutoff-independent} quantities such as mutual informations are independent of the specific Lagrangian presentation. This is consistent with the statement that a conformal field theory is defined by the spectrum of local operators and the operator product expansion.

We have shown the equivalence of the real-space entanglement spectra between boson and fermionic presentations at a specific point in the moduli space of $c = 1$ conformal theories, where both theories are free.  There is a line of conformal field theories, which corresponds to different radii of the free boson, and to a varying four-fermion coupling in the dual, modular-invariant fermionic theory (here we do not mean the ungauged Dirac fermion related by ``real duality").  We expect that the entanglement spectra of this whole line of theories are invariant under the bosonization map.

In contrast, one may consider the entanglement of regions in \emph{momentum} space, as discussed in \cite{Balasubramanian:2011wt}. This entanglement is an important aspect of  Wilsonian renormalization, in which ultraviolet degrees of freedom are integrated out, or traced over; in any interacting theory, the ultraviolet and infrared degrees of freedom are entangled in the ground state, and the state of the IR theory is described by a density matrix.  However, this entanglement is \emph{not} universal in the same sense; rather it depends very much on one's choice of presentation.  In the case of Bose-Fermi duality for arbitrary boson radius, if we choose the ultraviolet degrees of freedom to correspond to bosonic oscillators at high momentum, then the ground state will factorize between ultraviolet and infrared as the theory is free.  However, if we choose the UV degrees of freedom to correspond to fermionic oscillators at high momentum, then the four-fermion interactions guarantee that the ground state will be highly entangled between momentum scales, as is apparent by studying the explicit construction of the bosonic ground state in the interaction picture of the fermion theory \cite{Mattis:1964wp}.  In general, this presentation dependence is related to the fact that in calculations of real-space entanglement, the cutoff-dependent terms are scheme-dependent---they depend on the details of how one partitions the theory between IR and UV degrees of freedom. 

This is not to say that momentum-space entanglement is not useful; it is an important fact about integrating out UV degrees of freedom in an interacting theory \cite{Balasubramanian:2011wt}, and is a measure of the interactions of a given set of degrees of freedom.  But the real-space entanglement appear to be the right tool for characterizing theories in an invariant manner.

\subsubsection{Do R\'enyi entropies distinguish theories?}

Our study of the free fermion and self-dual boson has given evidence that entanglement entropies do indeed distinguish theories, if one includes regions with enough components. As we mentioned in Section \ref{sec:real_duality}, this question is related to an old program of Friedan and Shenker \cite{Friedan:1986ua}, who proposed that the set of partition functions on Riemann surfaces of all genera, as a function of the moduli of those Riemann surfaces, might completely characterize modular-invariant conformal field theories. The essential point is that the factorization limits build up these Riemann surfaces in terms of correlation functions on the sphere. The replica trick relates this proposal to the attempt to characterize conformal field theories via their R\'enyi entropies. The R\'enyi entropies, however, only depend on the Riemann surfaces on a slice through the full moduli space.  The question remains as to how much information about the conformal field theory can be extracted from this restricted class of partition functions. 

Finally, we would like to point out that there exists a large class of theories for which the distinguishing ability of entanglement entropies, along with the Friedan-Shenker program, fails badly in a specific limit. These are theories at large $c$ and strong coupling---very far from the free $c=1$ theories we've been considering so far. Specifically, any holographic CFT whose dual is Einstein gravity (possibly coupled to some matter) will have the same partition function at leading order in $1/c$ on a given Riemann surface. The reason is that the partition function is determined by the solution to the Euclidean Einstein equation whose boundary is the given Riemann surface; since none of the other fields are sourced by the boundary conditions, the solution will be locally AdS${}_3$ regardless of the matter content. In fact, we can go further: because the solution is locally AdS${}_3$, it will not be changed even in the presence of higher-derivative corrections to the bulk action, and the partition function will be changed only by an overall factor which amounts to a renormalization of the central charge. Such corrections correspond to moving away from infinite coupling in the boundary theory. Going even further, there is evidence that ``free" large-$c$ CFTs such as symmetric-product orbifolds also have the same partition functions as holographic ones \cite{Dijkgraaf:2000fq,Keller:2011xi}. (By ``large-$c$ theory" we mean one where the spectrum does not decompactify in the large-$c$ limit, i.e.\ the number of operators below any given scaling dimension remains finite.)

If all large-$c$ CFTs have the same partition functions on arbitrary Riemann surfaces, then they also have the same R\'enyi entropies for arbitrary $N$ and $n$. Further evidence for this proposition comes from several directions. First, the Ryu-Takayanagi formula \cite{Ryu:2006bv,Ryu:2006ef,Nishioka:2009un} gives the same results for the entanglement von Neumann entropies of any set of intervals in any theory whose ground state is represented by AdS${}_3$ (global or Poincar\'e), irrespective of what matter content the bulk theory might have. While this formula only applies when the bulk theory is Einstein gravity, it can be argued that this agreement survives higher-derivative corrections to the bulk action, based on the symmetries of AdS${}_3$ together with a standard ansatz for the effect of such corrections on the entropy \cite{Headrick:2010zt}. Finally, in \cite{Headrick:2010zt}, direct evidence was found using CFT techniques that the R\'enyi entropies are the same for all large-$c$ CFTs.

We have argued that a large class of theories have identical partition functions and entanglement spectra.\footnote{These arguments can be partially extended to higher-dimensional theories. Specifically, the arguments concerning holographic partition functions and the Ryu-Takayanagi formula can be extended to theories with duals controlled by Einstein gravity. On the other hand, the arguments concerning higher-derivative corrections rely on special properties of three-dimensional gravity and presumably do not extend to higher dimensions.} All of these arguments are approximate, in that they apply only to the leading (order-$c$) parts of the partition functions and entropies. It seems likely that $1/c$ corrections will indeed distinguish between theories.


\acknowledgments
We would like to thank Alejandra Castro, Vijay Balasubramanian, Dan Freedman, Matthias Gaberdiel, Chantal Hutchison, Daniel Jafferis, Matthew Kleban, Igor Klebanov, John McGreevy, Joe Polchinski, Massimo Porrati, Danny Ruberman, Eva Silverstein, Erik Tonni, Mark van Raamsdonk, Roberto Volpato, and Xi Yin for useful conversations.  M.H. and A.L. would like to thank the Kavli Institute for Theoretical Physics and the organizers of the ``Bits, Branes, and Black Holes" workshop for a stimulating environment while this work was in progress. M.H. would also like to Harvard University for hospitality while this work was being completed. M.H. is supported by the National Science Foundation under CAREER Grant No.\ PHY10-53842. A.L. is supported by DOE Grant DE-FG02-92ER40706. M.M.R. is supported by the Simons Postdoctoral Fellowship Program. This research was also supported in part by the National Science Foundation under Grant No.\ PHY11-25915.


\appendix

\section{A resummation of the self-dual boson partition function}\label{app:resummation}

In this appendix we will derive (\ref{eq:sd_lattice_high_genus}) from (\ref{eq:sdpfg}). Let $\Omega = A + i K$, with $A,K$ symmetric and real, and $K$ positive definite.  Plugging (\ref{eq:rsdef}) into the sum in (\ref{eq:sdpfg}),
\begin{eqnarray}
	&& I = \sum_{\vec\eps \in \Z_2^g}
		\big| \vartheta\left[ \begin{array}{c} \vec\eps \\ 0 \end{array} \right](0 | 2 \Omega)\big|^2
	= \sum_{\vec{m},\vec{n}\in \Z^g} \sum_{\vec\eps \in \Z_2^g}
		\exp\left[ 2\pi i (n + \eps)^i A_{ij} (n + \eps)^j  \right.\nonumber\\
		& & \ \ \ \ \ \ \ \ \ \ \ \ \ \left. - 2\pi i (m+\eps)^i A_{ij}(m+\eps)^j - 2\pi (n+\eps)^i K_{ij}(n+\eps)^j - 2\pi
			(m+\eps)^i K_{ij}(m+\eps)^j\right]\nonumber\\
			 & & \label{eq:sdtheta}
\end{eqnarray}
Using 
\be
	2x^i K_{ij}x^j + 2 y^i K_{ij}y^j = (x-y)^i K_{ij}(x-y)^j + (x+y)^i K_{ij} (x+y)^j,\label{eq:sumdiff}
\ee
the sum in (\ref{eq:sdtheta}) can be rewritten as:
\begin{eqnarray}
	& & I = \sum_{\vec{m},\vec{n}\in \Z^g} \sum_{\vec\eps \in \Z_2^g}
	\exp\left[ 2\pi i (n + \eps)^i A_{ij} (n + \eps)^j - 2\pi i (m+\eps)^i A_{ij}(m+\eps)^j\right. \nonumber\\
	& & \ \ \ \ \ \ \ \ \left.- \pi (n + m + 2\eps)^i K_{ij} (n + m + 2\eps)^j - \pi (n-m)^i K_{ij} (n-m)^j\right]\label{eq:sdtwo}
\end{eqnarray}
Defining ${\vec \ell} = ({\vec n} - {\vec m})$, we find $\vec{n} +{\vec m} + 2{\vec \eps}  = {\vec \ell} + 2 {\vec m} + 2\eps$. Eq. (\ref{eq:sdtwo}) can now be rewritten as:
\begin{eqnarray}
	& & I = \sum_{\vec{m},\vec{\ell}\in \Z^g} \sum_{\vec\eps \in \Z_2^g}
	\exp\left[ \half \pi i (2\ell  + 2 m + 2 \eps)^i A_{ij} (2\ell + 2 m + \eps)^j - \half \pi i (2 m+ 2\eps)^i A_{ij}(2 m+2 \eps)^j\right. \nonumber\\
	& & \ \ \ \ \ \ \ \ \left.- \pi (\ell + 2m + 2\eps)^i K_{ij} (\ell + 2m + 2\eps)^j - \pi\ell^i K_{ij} \ell^j\right]\label{eq:sdthree}
\end{eqnarray}
$m$ appears only in the form $2m + 2\eps$.  Summing over all $m$ means summing over even $2m$; since $\vec \epsilon \in \{0,\half\}^g$, summing over $m$ and $\eps$ means $2{\vec m} + 2{\vec \eps}$ take all values in $\Z^g$ once; we can thus replace the sums over ${\vec m}$ and  ${\vec{\eps}}$ with a sum over ${\vec k} = 2{\vec m} + 2{\vec \eps}$, so that (\ref{eq:sdthree}) becomes
\be\label{eq:sdfour}
	I = \sum_{\vec{\ell},\vec{k}} \exp\left[\pi i( 2\ell + k)^i A_{ij} (2\ell + k)^j - \pi i k^i A_{ij} k^j - \pi (\ell + k)^i K_{ij} (\ell + k)^j - \pi \ell^i K_{ij} \ell^j \right]
\ee
Shifting ${\vec k} \to \vec{k} - {\vec\ell}$, (\ref{eq:sdfour}) becomes
\be
	I = \sum_{\vec{\ell},\vec{k}} \exp\left[2\pi i\ell^i A_{ij} k^j  - \pi \ell^i K_{ij} \ell^j - \pi k^i K_{ij} k^j \right],\label{eq:resummation_result}
\ee
and recalling that the self-dual momentum lattice is simply $ k_L \pm k_R \in \sqrt{2}\Z$, we recover (\ref{eq:sd_lattice_high_genus}).

\section{Representations of the dihedral group}\label{app:dihedral_rep}

The dihedral group $D_n$ is generated by two elements: $g$ such that $g^n = 1$, generating a $\Z_n$ subgroup, and $\tau$ such that $\tau^2 = 1$, which satisfy the relation $g\tau = \tau g^{-1}$. A good reference for this subject is \cite{Lomont:1959}.

For $n$ odd there are $\frac{n+3}{2}$ irreducible representations:
\begin{itemize}
\item The trivial representation ${\bf 1}$.  
\item An additional one-dimensional representation ${\bf T}$ with $g = 1$, $\tau = -1$.
\item $\frac{n-1}{2}$ two-dimensional representations $R_{k = 1,\ldots,\frac{n-1}{2}}$ with
\be
	g = \left(\begin{array}{cc} e^{\frac{2\pi i k}{n}} & 0 \\ 0 & e^{\frac{-2\pi i k}{n}} \end{array}\right)\ ;\ \ \ \ \ 
	\tau = \left(\begin{array}{cc} 0 & 1 \\ 1 & 0 \end{array}\right)
\ee
One may write these representations for any $k$, but for $k > (n-1)/2$, they are equivalent to one of the above.
\end{itemize}

For $n$ even there are $\frac{n + 6}{2}$ irreducible representations:
\begin{itemize}
\item The trivial representation ${\bf 1}$.
\item The one-dimensional representation ${\bf T}$ with $g = 1$,$\tau = -1$.
\item The one-dimensional representation ${\bf S}$ with $g = -1$,$\tau = 1$.
\item The one-dimensional representation ${\bf U}$ with $g = -1$,$\tau = -1$.
\item $\frac{n-2}{2}$ two-dimensional representations $R_{k = 1,\ldots,\frac{n-1}{2}}$ with
\be
	g = \left(\begin{array}{cc} e^{\frac{2\pi i k}{n}} & 0 \\ 0 & e^{\frac{-2\pi i k}{n}} \end{array}\right)\ ;\ \ \ \ \ 
	\tau = \left(\begin{array}{cc} 0 & 1 \\ 1 & 0 \end{array}\right)
\ee
One may write these representations for any $k$, but for $k > (n-2)/2$, they are equivalent to one of the above.
\end{itemize}

\section{The boson at arbitrary radius}\label{app:genl_boson_renyi}

Using the replica trick and classic results on correlators of twist fields in orbifold CFTs, Calabrese, Cardy, and Tonni (CCT) \cite{Calabrese:2009ez} calculated the R\'enyi entropies $S_n^R(A)$ for two intervals for the compact boson at arbitrary radius $R$; their result is shown in \eqref{CCT}. In this appendix we will use the techniques developed in Section 3 to reproduce their result in a different way.

The basic idea is do a direct calculation of the partition function on the replicated surface $\Sigma_{n,2}$ by Weyl-transforming it to a non-singular surface and then applying \eqref{eq:sdpfg} (or more precisely its generalization to arbitrary radius). However, to avoid having to compute the Liouville action \eqref{Liouville} and the Laplacian determinant appearing in \eqref{eq:sdpfg}, we consider a ratio of the partition function of the boson to that of the Dirac fermion, and use Casini, Fosco, and Huerta's result \eqref{CasiniHuerta} for the latter \cite{Casini:2005rm,Casini:2009vk}. All of the complicating factors are the same for the two theories, since they have the same central charge and same oscillators, and therefore cancel in the ratio, leading to quite a simple calculation. Furthermore, this derivation explains why the boson result takes the form of the fermion result plus a correction term.

Let us proceed with the calculation. Since both theories have $c=1$, they have the same Liouville action for any Weyl transformation; therefore the ratio of partition functions is independent of the choice of fiducial metric:
\begin{equation}\label{bootstrap}
S_n^R(A) = S_n^{\rm (f)}(A) + \frac1{1-n}\ln\frac{Z^R}{Z^{\rm (f)}} = S_n^{\rm (f)}(A) + \frac1{1-n}\ln\frac{\hat Z^R}{\hat Z^{\rm (f)}}\,.
\end{equation}
The partition function for the boson at arbitrary $R$ is essentially given by the same formula as in the self-dual case \eqref{eq:sd_lattice_high_genus}, but with the momentum lattice $\Gamma^{\rm (b)}$ (defined in \eqref{Gammabdef}) generalized to
\begin{equation}
\Gamma^R = \left\{(k_L,k_R):k_L+k_R\in\sqrt{2/\eta}\Z,k_L-k_R\in\sqrt{2\eta}\Z\right\}
\end{equation}
(recall that $\eta = R^2/R_{\rm sd}^2$). It's useful to express $k_{L,R}$ in terms of the integral momentum and winding numbers $n,w$:
\begin{equation}
k_{L,R} = \frac1{\sqrt{2\eta}}n\pm\sqrt{\frac\eta2}w\,.
\end{equation}
The dot product appearing in the partition function becomes
\begin{equation}
k\cdot k' = k_Lk_L'+k_Rk_R' = \frac1{\eta}nn'+\eta ww'\,.
\end{equation}
As we showed in subsection 3.4, for $N=2$, the period matrix is always imaginary ($A=0$, $\Omega=iK$). Hence the ratio of partition functions is
\begin{equation}\label{sofar}
\frac{\hat Z^R}{\hat Z^{\rm (f)}} = 2^{(1-g)/2}\frac{\vartheta(0|\Omega/\eta)\vartheta(0|\eta\Omega)}{\vartheta(0|\Omega)^2}\,.
\end{equation}
As discussed at the end of subsection 3.3, the factor of $2^{(1-g)/2}$ contributes a constant to the non-universal part of the R\'enyi entropies, and can be neglected. We thus have
\begin{equation}\label{CCT2}
S_n^R(A) = S_n^{\rm (f)}(A) + \frac1{1-n}\ln\frac{\vartheta(0|\Omega/\eta)\vartheta(0|\eta\Omega)}{\vartheta(0|\Omega)^2}\,.
\end{equation}
This result is extremely similar to the CCT result \eqref{CCT}, the only difference being that the matrix $\Gamma$ appearing there is replaced by the period matrix $\Omega$. However, CCT claim based on numerical evidence that for any $\eta$,
\begin{equation}
\vartheta(0|\eta\Omega) = \vartheta(0|\eta\Gamma)
\end{equation}
(see appendices A, B, C of \cite{Calabrese:2009ez}), establishing that \eqref{CCT2} and \eqref{CCT} agree. In fact, CCT give an explicit formula for $\Omega$, which is no more complicated than that for $\Gamma$, so \eqref{CCT2} may itself be useful for direct calculation of the R\'enyis.

Note that the result \eqref{CCT2} holds not just for the case $N=2$, arbitrary $n$, treated by CCT, but also for the case $n=2$, arbitrary $N$, since $\Omega$ is imaginary there as well (as shown in subsection 3.4). However, to apply \eqref{CCT2}, one would need to compute $\Omega$. The Riemann surface $\Sigma_{2,N}$ is hyperelliptic in that case, and one might be able to compute its period matrix via the associated Picard-Fuchs equations.

\bibliographystyle{JHEP}
\bibliography{entanglerefs}

\end{document}